\documentclass[pre,preprint,showpacs]{revtex4}
\usepackage{graphicx}
\usepackage{amsmath}
\usepackage{setspace} 

\def\eq#1{Eq.~(\ref{#1})}
\def\fig#1{Fig.~\ref{#1}}
\def\tab#1{Tab.~\ref{#1}}
\def\sec#1{Sec.~\ref{#1}}

\newcommand{\boldv}[1]{{\mathbf #1}}

\newcommand{\ha}{\hspace{1cm}}

\newcommand{\grad}{{\mathbf \nabla}}

\newcommand{\tr}{\mathrm{tr}}
\newcommand{\mean}[1]{\langle #1 \rangle}

\newcommand{\R}{R}    
\newcommand{\Rr}{R}   
\newcommand{\rc}{x_c} 
\newcommand{\kk}{\kappa} 

\begin{document}

\title{Dynamic states of cells adhering in shear flow:\\
from slipping to rolling}

\author{C. B. Korn}
\author{U. S. Schwarz}

\affiliation{University of Heidelberg, Bioquant, BQ 0013 BIOMS Schwarz,
Im Neuenheimer Feld 267, D-69120 Heidelberg, Germany}

\begin{abstract}
  Motivated by rolling adhesion of white blood cells in the vasculature, we
  study how cells move in linear shear flow above a wall to which they can
  adhere via specific receptor-ligand bonds.  Our computer simulations are
  based on a Langevin equation accounting for hydrodynamic interactions,
  thermal fluctuations and adhesive interactions. In contrast to earlier
  approaches, our model not only includes stochastic rules for the formation
  and rupture of bonds, but also fully resolves both receptor and ligand
  positions. We identify five different dynamic states of motion in regard to
  the translational and angular velocities of the cell. The transitions
  between the different states are mapped out in a dynamic state diagram as a
  function of the rates for bond formation and rupture.  For example, as the
  cell starts to adhere under the action of bonds, its translational and
  angular velocities become synchronized and the dynamic state changes from
  slipping to rolling. We also investigate the effect of non-molecular
  parameters. In particular, we find that an increase in viscosity of the
  medium leads to a characteristic expansion of the region of stable rolling
  to the expense of the region of firm adhesion, but not to the expense of the
  regions of free or transient motion. Our results can be used in an inverse
  approach to determine single bond parameters from flow chamber data on
  rolling adhesion.
\end{abstract}


\maketitle

\section{Introduction}

Rolling adhesion of white blood cells (leukocytes) along the vascular
endothelium plays a key role in the immune response and is a prominent example
for the interplay between transport and specific adhesion in biological
systems \cite{c:spri94,alon:97a,alon:07b}. Random encounters between cell and
vessel wall lead to the formation of initial bonds based on adhesion receptors
from the selectin-family.  Because selectins are characterized by fast
association and dissociation kinetics, new bonds are readily formed on the
downstream side and old bonds are continuously ruptured at the upstream side.
If the processes of bond formation and rupture are sufficiently balanced,
rolling adhesion results. As rolling velocity is smaller than the velocity of
a cell moving freely in hydrodynamic flow, this mechanism allows the
leukocytes to more efficiently survey the vessel wall for appropriate
molecular signals.  The main signals in this context are chemokine molecules,
which indicate the presence of an infection and lead to firm adhesion based on
the adhesion receptors from the integrin-family. Firm adhesion in turn is a
prerequisite for transendothelial migration into the surrounding
tissue. Similar mechanisms are used by stem and cancer cells to disseminate in
the body through the blood flow.

The main tool for investigating rolling adhesion under controlled
conditions are flow chambers \cite{lawrence:91}. There
receptor-carrying cells suspended in an aqueous solution are perfused
in linear shear flow above a ligand-coated wall. For example, it was
shown with flow chamber experiments that selectin bonds have fast
kinetic rates and that the dissociation rate increases with force
(slip bond), thus explaining their superior function in mediating
rolling adhesion \cite{alon:95,alon:97}. Recently it was demonstrated
by a combination of atomic force microscopy and flow chamber
experiments that the lifetime of single selectin bonds shows a
biphasic response under force \cite{marshall:03,mcever:06}. Such a
catch-slip bond behavior might have evolved to avoid adhesion of
leukocytes under static conditions. Indeed it has been found early in
flow chamber experiments that leukocytes adhere only above a critical
threshold of shear \cite{finger:96}. In addition to specific
properties of the molecular bonds, transport processes might also play
an important role in creating the shear threshold
\cite{schwarz:04,yago:07}.  Both rolling adhesion \cite{hammer:96a}
and the shear threshold \cite{hammer:00a} have also been demonstrated
in cell-free flow chamber experiments with ligand-coated micron-sized
beads. Although cell-free rolling is more erratic than leukocyte
rolling, this indicates that cellular features (like cell
deformability) are not essential to rolling adhesion.  Flow chamber
experiments also allow to study the exact effect of non-molecular
parameters, for example of changes in the viscosity of the medium. By
adding an inert substance like the sugar ficoll, it is possible to
change shear stress but not shear rate in a flow chamber experiment,
thus dissecting the respective roles of force and transport
\cite{springer:01,schwarz:04,yago:07}.

The standard observable in flow chamber experiments is the translational
velocity as a function of time \cite{alon:97,yago:04,alon:05}.  From this time
series further variables can be derived, for example average and standard
deviation of velocity. In the case of low ligand density, a stopped cell is
most likely held by a single bond and detaches without rebinding. Then the
time series can be used to measure single bond lifetime. At higher ligand
density, rolling occurs. Because rolling is never smooth but erratic due to
the stochastic processes on the molecular level, then the time series can be
used to measure pause time distributions.  The velocity time series can also
be used to define different states of motion of the cell. Typically a cell is
said to undergo rolling whenever its mean velocity (averaged over some
seconds) significantly decreases compared to the free (hydrodynamic)
velocity. If no motion can be detected for several seconds, the cell is
considered to be in firm adhesion \cite{alon:05}.

Because rolling adhesion is not only of large physiological importance, but
also characterized by an intricate interplay of different physical factors, it
has long been subject to intense modeling efforts. If molecular effects are of
interest, then such modeling efforts typically start with physical models for
bond association and dissociation \cite{bell:78}. Combined with the
hydrodynamics of a sphere in front of a wall, they lead to algorithms known as
adhesive dynamics \cite{hammer:92}.  This approach has been applied before to
different aspects of rolling adhesion, e.\,g., the interplay of two receptor
systems \cite{hammer:03} or the effect of catch bonds
\cite{hammer:07a}. Recently we have developed a new variant of this algorithm
which in contrast to earlier approaches fully resolves the spatial positions
of the receptors on the sphere and the ligands on the wall.  Using this
approach, we were able to predict the efficiency of initiating cell adhesion
in shear flow as a function of the density and geometry of the receptor and
ligand patches \cite{korn:06,korn:07a}. A large modeling effort has also been
spent on the role of cell deformability
\cite{pozrikidis:03book,tran:03,truskey:05,jadhav:05,balazs:06} and the
interaction between multiple particles \cite{munn:05,king:01a,balazs:06}.
Here we focus on the case of moderate shear flow and small numbers of
cells, when cell deformability and hydrodynamic interactions between
cells are not relevant \cite{korn:07a}.

A convenient way to present the results from adhesive dynamics
simulations of rolling adhesion is the calculation of state diagrams
which predict different types of motion over a large range of model
parameters \cite{hammer:00,hammer:05}. Because in flow chamber
experiments one usually measures only translational velocity as a
function of time, these state diagrams have been determined before
based only on the translational velocity simulated as a function of
different molecular parameters. However, computer simulations also
allow to track the angular velocity of a cell, thus doubling the
number of degrees of freedom that could be compared. For example, in
physical terms cell rolling means that translation and rotation are
synchronized. Although this fact was already noted in the pioneering
paper on adhesive dynamics \cite{hammer:92}, it has not been
systematically exploited due to the lack of experimental data. Here we
present a detailed analysis of the different dynamic states of rolling
adhesion which is based on the simultaneous tracking of both
translational and rotational degrees of freedom. In contrast to
earlier work in this field, we fully resolve the spatial positions of
receptors and ligands and calculate the state diagrams as a function
of the on- and off-rate. Moreover, we explore the role of different
external parameters, including the viscosity of the medium. 

The paper is organized as follows. In \sec{sec:algorithm} we explain
our numerical method, which is a combination of a Langevin equation
accounting for hydrodynamic and thermal forces in the limit of small
Reynolds numbers (Stokesian dynamics) \cite{brady:89} and adhesive
dynamics accounting for the forces caused by the formation of adhesive
bonds between cell and substrate \cite{hammer:92}.  By taking care of
the diffusive motion of the sphere, we are able to explicitly resolve
both receptor and ligand positions. As a first application of the
hydrodynamic part of our model, we explain the physical difference
between slipping and rolling. We close this section with a detailed
description of the parameters relevant for the algorithm. In
\sec{sec:def_states} we point out, using a simple analytical
description, how the cell slows down and simultaneously synchronizes
its translational and angular velocity under the action of bonds. Then
we discuss the mean velocities of the cell as a function of the on-
and off-rate of the receptor-ligand bonds. Based on this we define
five distinct states of stationary motion for a cell in a flow
chamber.  In \sec{sec:generalstate} we identify these different states
in state diagrams displaying the dependence on the internal bond
parameters, i.\,e., the on- and off-rate. In addition, the effect of
external parameters to the location of the states in the space of
different on- and off-rates is discussed. In the closing section,
\sec{sec:discussion}, we argue that molecular parameters can be much
better extracted from flow chamber experiments if both translational
and rotational degrees of freedom are measured, as it is done in our
simulations. Furthermore we propose experiments based on recent
nanotechnological developments that could result in flow chamber data
for the angular velocities of cells or microspheres.

\section{Model and simulation method}
\label{sec:algorithm}

\subsection{Stokesian dynamics}
\label{sec:algo:stokes}

\begin{figure}
  \begin{center}
      \resizebox{.8\linewidth}{!}{\includegraphics{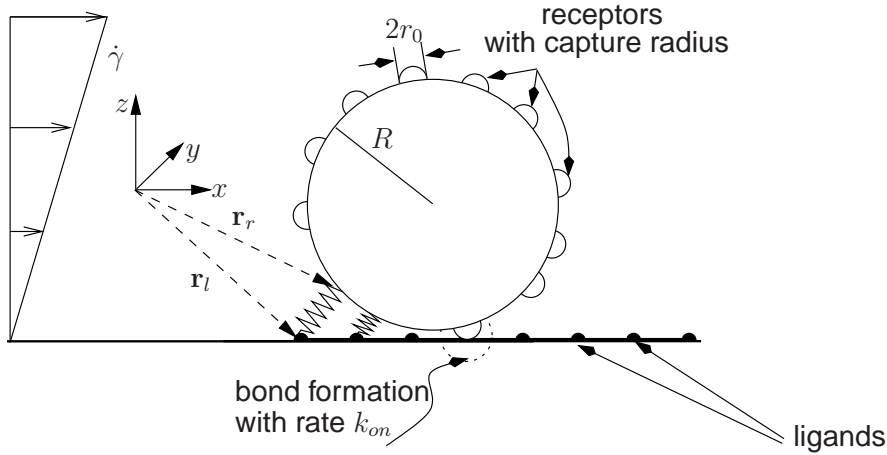}}
    \caption{Model system and adhesive dynamics. A rigid sphere of radius
    $R$ moves above a wall. The shear flow is linear with shear rate $\dot \gamma$.
    Receptors are modeled as sticky spheres with capture radius $r_0\ll R$ on the
    surface of the sphere. Ligands are modeled as dots on the boundary wall. A receptor-ligand
    bond forms with rate $k_{on}$ whenever a receptor patch has some overlap
    with a ligand. Bonds are modeled as Hookean springs and
    exert a force on the sphere that is proportional to the distance between
    the receptor position $\boldv{r}_r$ and the ligand position
    $\boldv{r}_l$. They rupture with a force-dependent rate $k_{off}(F)$.
     \label{fig:model:bdynamics}
   }
\end{center}
\end{figure}

As a simple model system for a cell or a microsphere in a flow chamber we use
a rigid sphere of radius $R$ that moves above a wall
(cf. \fig{fig:model:bdynamics}). For objects as small as white blood cells
(with diameters $<10\ \mu$m) the typical value for the Reynolds number is much
less than one and inertia can be neglected (overdamped motion). Therefore, the
flow around the cell is laminar and well described by the linear Stokes
equation.
Cells are usually observed only close to the wall, that is at a distance to
the wall which is much smaller than the separation between the two parallel
glass plates used in a flow chamber. In this region the unperturbed flow
profile is approximately linear. Throughout this paper we therefore consider
the sphere to flow in simple shear flow, i.\,e., we consider the shear rate
$\dot\gamma$ to be a constant.

In the situation under consideration the sphere is not only driven by
the imposed shear flow but also by direct forces exerted on the cell.
These include a constant gravitational drift force which results from
a slight density difference between the cells and the surrounding
medium.  For flow chamber experiments, this implies that most cells
adhere to the bottom plate, which usually is also the plate coated
with ligands.  In our context, the strongest forces result after
initial binding from receptor-ligand bonds that pull on the sphere. A
third source of force are thermal forces which are ubiquitous for
objects in the micrometer range and lead to diffusive motion.  This
kind of motion can be described by an appropriate Langevin equation.
It is convenient to introduce a six-dimensional state vector
$\boldv{X}$ which includes both the three translational and the three
rotational degrees of freedom. The first three components of
$\boldv{X}$ denote the Cartesian coordinates with respect to a flow
chamber based coordinate system (cf. \fig{fig:model:bdynamics}). The
last three components describe the rotation that maps a coordinate
system fixed to the center and the orientation of the sphere to the
flow chamber fixed coordinate system.  Similar compact notations are
introduced for velocities and forces.  The symbol $\boldv{U}$ denotes
a six-dimensional velocity with the first three components being
translational velocities and the last three being angular velocities.
$\boldv{F}$ is the combined force and torque vector. Using this notation,
the Langevin equation describing the motion of the cell reads
\cite{korn:07a}
\begin{equation}
  \label{langevin-ito}
  \dot{\boldv{X}} = \boldv{U}^\infty + \mathsf{M}(\boldv{F}^S + \boldv{F}^D )
  + k_B T\grad\mathsf{M} + \boldv{g}_t^I.
\end{equation}
Here, $\mathsf{M}$ is the $6\times 6$ mobility matrix.  $\boldv{U}^\infty
\propto \dot\gamma$ is the velocity of the unperturbed shear flow at the
position of the center of the sphere.  $\boldv{F}^S\propto\dot\gamma$ is the
shear force, which results from the hydrodynamic interaction between the cell
and the wall. $\boldv{F}^D$ denotes all direct forces (torques) acting on the
sphere, like gravity and bond forces. The terms $k_B T\grad\mathsf{M}$ and
$\boldv{g}_t^I$ describe the effect of thermal noise.  $\boldv{g}_t^I$ is
Gaussian white noise with
\begin{align}
  \label{noise}
  \mean{\boldv{g}^I_t} = 0, \ha \mean{\boldv{g}^I_{t'}\boldv{g}^I_t} = 2 k_B T_a
  \mathsf{M} \delta(t-t')\ .
\end{align}
Here $T_a$ is the ambient temperature and $k_B$ Boltzmann's constant. The
mobility matrix for a sphere above a wall depends the distance between sphere
and wall \cite{jones:98}. Thus, the noise is multiplicative, leading to the
gradient term $\grad\mathsf{M}$ in \eq{langevin-ito}. We interpret the noise
$\boldv{g}_t$ in the usual Stratonovich sense. However, \eq{langevin-ito} is
written in its It{\^o} version \cite{honerkamp:94,gardiner:85}, as indicated
by the superscript $I$.  This allows to directly derive a simple Euler
algorithm for the update step $\Delta\boldv{X}$ of the configuration of the
sphere during a time step $\Delta t$. In non-dimensional form, using the
radius $R$ of the sphere as the length scale, the inverse shear rate
$1/\dot\gamma$ as the time scale, and $6\pi\eta R^2\dot\gamma$ as the force
scale, the first order discretized version of \eq{langevin-ito} reads
\begin{align}
  \label{langevin-euler-units}
  \Delta \boldv{X}_t = (\boldv{U}^\infty +
  \mathsf{M}(\boldv{F}^S + {\boldv{F}}))
  \Delta t +  \frac{1}{Pe}\grad \mathsf{M}\Delta t + 
  \sqrt{\frac{1}{Pe}} \boldv{g}(\Delta t) + \mathcal{O}(\Delta t^2).
\end{align}
Here the dimensionless P{\'e}clet number $Pe = (6\pi\eta R^3\dot\gamma)/(k_B
T_a)$ describes the relative importance of convective versus diffusive motion
of the sphere.  The first two moments for the Gaussian white noise now read
\begin{align}
  \label{disc_noise}
  \mean{\boldv{g}(\Delta t)} = 0, \ha \mean{\boldv{g}(\Delta t)
  \boldv{g}(\Delta t)} = 
  2\mathsf{M}\Delta t\ .
\end{align}
The algorithm \eq{langevin-euler-units} is also known as Stokesian dynamics
and has been derived, e.\,g., by Brady and Bossis \cite{brady:89} and for
vanishing shear flow by Ermak and McCammon \cite{ermak:78}. A more detailed
description of our algorithm is given in Refs.~\cite{korn:07a,korn:phd}. In
order to obtain accurate results when simulating the motion of the sphere it
is essential to properly calculate the mobility matrix $\mathsf{M}$ and the
shear force $\boldv{F}^S$. In our simulations we use a numerical method
described by Jones et al. \cite{jones:92,jones:98} which allows to accurately
calculate the components of the mobility matrix for a single sphere above a
wall for arbitrary sphere wall distances. The values obtained by this method
agree very well with the classical results given for some tabulated height
values by Goldman et al. \cite{goldman:67b}.

From \eq{langevin-euler-units} one readily sees that diffusive motion is less
relevant if the P{\'e}clet number $Pe$ is large. Using typical values for
white blood cells (cf. \tab{tab:leuk:parameters}) one finds that $Pe$ is of
the order $10^4$. Therefore, regarding the motion in flow direction thermal
motion can be neglected on the scale of the cell radius. However, for the
motion in $z$-direction the only force on a freely flowing cell is
gravitation, which is small due the small  density difference between the cell
and the surrounding medium and therefore comparable to thermal forces in this
direction \cite{korn:07a}.  For this reason, in our work we use the full
Langevin equation from \eq{langevin-ito}. Another reason why we keep the
diffusion terms in \eq{langevin-euler-units} despite the large $Pe$ values has
to do with the size of the receptors on the surface of the cell and will
become more apparent in the next subsection, \sec{sec:algo:addyn}.

\begin{figure}[t!]
  \begin{center}
    \includegraphics[width=.78\linewidth]{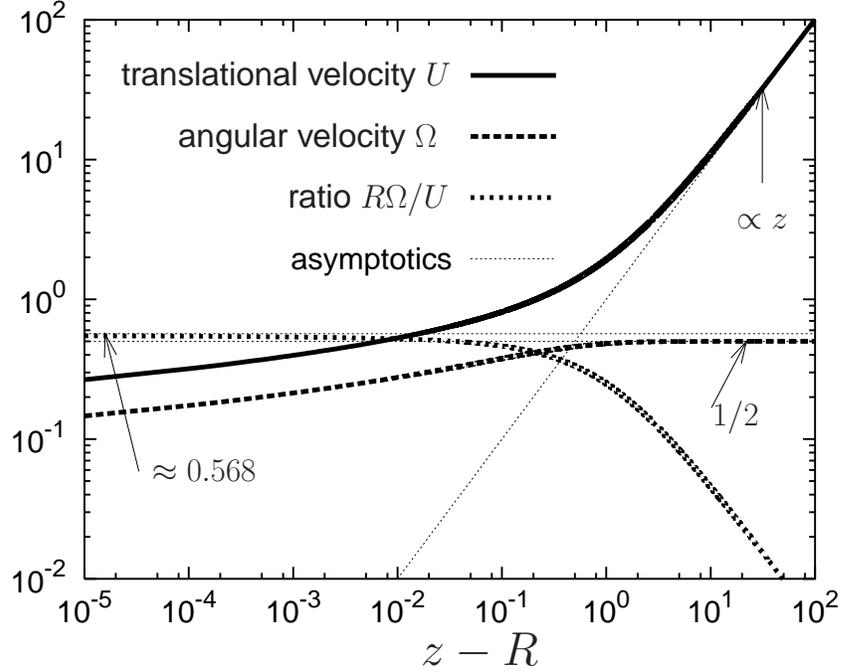}
    \caption{Velocities of the sphere in the limit of deterministic motion,
      $Pe \rightarrow \infty$.  Translational and angular velocity $U$ and
      $\Omega$, respectively, and their ratio $R\Omega/U$ are shown as a
      function of the separation $z - R$ between sphere and wall. The thin
      lines depict the asymptotic behavior for $z \rightarrow R$ and $z
      \rightarrow \infty$, respectively. $U$ and $\Omega$ are plotted in units
      of $R\dot\gamma$ and $\dot\gamma$, respectively. $z$ is measured in
      units of $R$.}
      \label{fig:model:velocities}
  \end{center}
\end{figure}

For the following it is important to consider how the translational
velocity $U$ (in $x$-direction) and the angular velocity $\Omega$ (for
rotations about the $y$-axis) behave in the limits of small and large
separation from the wall. If the translational and angular velocities
were completely synchronized, the ratio $R\Omega/U$ should be equal to
unity. This would correspond to rolling in a macroscopic sense, e.g.\
for a sticky sphere rolling down an inclined plane or for a car wheel
on the street.  We now show that the situation is different for a cell
in free hydrodynamic motion. For $z$ denoting the height of the
center of the sphere above the wall, $z-R$ denotes the gap between
sphere and wall. In \fig{fig:model:velocities} we plot $U$ and
$\Omega$ as a function of $z-R$ in the limit of deterministic motion,
$Pe \rightarrow \infty$.
For the sphere far away from the wall, $z - R\rightarrow \infty$, we
have the case of free flow, thus $U = \dot \gamma z$ and $\Omega =
1/2$. Therefore the ratio $R\Omega/U$ vanishes in this limit.  For the
sphere approaching the wall, $z - R \rightarrow 0$, both $U$ and
$\Omega$ slowly approach zero due to the hydrodynamic no-slip boundary
condition.  The ratio $R\Omega/U$ however approaches a finite limit
$R\Omega/U \approx 0.5676$ as computed by Goldman et al.
\cite{goldman:67b}. Together these results show that hydrodynamic
interactions increasingly synchronize translational and angular
velocities as the cell approaches the wall. However, the ratio
$R\Omega/U$ never reaches the value $1$ as it would for rolling in a
macroscopic sense.  Therefore, the hydrodynamic coupling between the
cell and the wall is not strong enough to lead to rolling and a freely
moving cell is always slipping.

\subsection{Bond dynamics}
\label{sec:algo:addyn}

We now include receptors, ligands and receptor-ligand bonds into our model.
Throughout this paper we call the cellular part of a bond \emph{receptor} and its
counter part on the wall \emph{ligand}. Before a cell receptor can react with a wall
ligand to form a complex (bond), there must be a physical transport process
which brings the two components to close proximity. Formally bond formation
can be separated into a transport and a reaction step using the notion of an
encounter complex \cite{berg:77}.  The encounter complex is formed whenever a
receptor and a ligand are less than the capture radius $r_0$ away from each
other.  Therefore, we model receptors as \emph{reactive patches} on the sphere
surface having a spherical capture range of radius $r_0$. The capture length
$r_0$ bridges the gap between the continuum approach followed by implementing
linear hydrodynamics and the discrete nature of the receptor
molecules. Ligands are modeled as dots on the boundary wall. Because we
explicitly resolve receptor and ligand positions, it is essential to also
include diffusive motion in the algorithm, \eq{langevin-euler-units}.
Although diffusion plays a minor role for the relative position of the cell,
it affects the positions of its surface receptors, which are of much smaller
size than the sphere itself.

During the time an encounter complex exists it can react to the final
receptor-ligand bond complex with the on-rate $k_{on}$. Reversely, any bond
complex can rupture into the encounter state with the off-rate
$k_{off}$. Experimentally it has been found that the dissociation rate depends
on the physical force acting on the bond complex \cite{alon:95,alon:97}. These
early experiments agreed nicely with the simplest model for bond dissociation
under force, which had been conceived first by Bell \cite{bell:78}:
\begin{align}
  \label{eq:model:bell}
  k_{off}(F) = k_{0}\exp(F/F_d),
\end{align}
with $k_0$ the dissociation rate at zero pulling force, $F$ the force on the
bond and $F_d$ the detachment force scale.  The Bell model,
\eq{eq:model:bell}, can be rationalized using Kramers theory as being a
thermally activated escape over a transition state barrier in the presence of
an external force \cite{kramers:40,shillcock:98,evans:97,evans:07}.  Although recent
evidence suggests that the dissociation rate shows a more complicated force
dependence at small forces \cite{marshall:03,mcever:06}, \eq{eq:model:bell}
has been demonstrated to properly describe the dissociation process for
selectin bonds in the high force regime \cite{springer:01}. Here we use the
Bell model for conceptual and computational simplicity.

In order to include the probabilistic nature of bond formation and rupture,
the algorithm \eq{langevin-euler-units} has to be extended to include rules
that take care of these processes. Such rules have been first set-up by Hammer
and co-workers \cite{hammer:92} and been refined many times to model various
aspects of leukocyte rolling. They are now known as \emph{adhesive dynamics}.
Here, we briefly explain the main idea behind these rules. A detailed
description of our implementation of these rules is then given in the
appendix, \sec{appendix:addyn}.  In order to simulate the motion of a cell
under the action of bonds, the configuration of the sphere is updated at each
time step $\Delta t$ according to \eq{langevin-euler-units}. If, after some
update, a receptor patch on the surface of the sphere overlaps with a wall
ligand a receptor-ligand bond forms with rate $k_{on}$, i.\,e., the
probability for bond formation during time step $\Delta t$ is $1 -
\exp(-k_{on}\Delta t)$. Existing bonds are modeled as harmonic springs,
i.\,e., the force along these bonds is proportional to the bond
extension, essentially given by $\|\boldv{r}_r - \boldv{r}_l\|$, with
$\boldv{r}_r$ and $\boldv{r}_l$ being the receptor and ligand position,
respectively (see \fig{fig:model:bdynamics}). Force and torque resulting from
extended bonds enter then the configuration update equation,
\eq{langevin-euler-units}, via the direct force term $\boldv{F}^D$.  Similarly
to bond formation, the probability for bond rupture during time step $\Delta
t$ is given by $1 - \exp(-k_{off}\Delta t)$, where $k_{off}$ is the force
dependent dissociation rate according to \eq{eq:model:bell}.

In contrast to earlier implementations of adhesive dynamics, we explicitly
resolve both receptors and ligands in space, rather than considering a wall
homogeneously coated with ligands at constant density. One immediate advantage
of our method is that it avoids a flow rate-dependent rate of bond formation
\cite{hammer:99}. In order to be able to spatially resolve both ligands and
receptors, it is necessary to include the Brownian motion of the cell.  If one
considered only the deterministic part of \eq{langevin-ito}, at low densities
of receptors and ligands it could happen that a receptor never finds a
ligand. Although these extensions of adhesive dynamics lead to increased
computational effort, they are closer to real biological systems, which on the
molecular level are in permanent thermal motion. In particular, in the future
our implementation will allow us to model receptor-ligand kinetics in more
molecular detail.  As will be discussed later, our algorithm also opens up the
perspective to compare adhesive dynamics simulations to flow chamber
experiments using substrates with nanopatterned ligand.

\subsection{Parameters}
\label{sec:param}

Our model contains thirteen different dimensional para\-meters.  With
$R$, $1/\dot\gamma$, $6\pi\eta R^2 \dot\gamma$ being the natural
scales of length, time and force, respectively, we are left with ten
numerical (dimensionless) parameters appearing in the algorithm.
Typical values for these (both the dimensional and dimensionless)
parameters are listed in \tab{tab:leuk:parameters}.  The parameters
$R, T_a, \dot \gamma, \eta, \Delta \rho$ influence the flow properties
and besides viscosity we keep these parameters fixed.  For the ambient
temperature $T_a$ we choose room temperature $T_a = 293$~K. For flow
chamber experiments with cells, the choice $T_a = 310$~K would be more
appropriate, but the exact choice of this value, which affects bond
dissociation kinetics and effective strength of diffusion, turns out
to be irrelevant for the physiologically relevant parameter values
chosen here. For the Stokes radius $R$ of the cells we use $R =
4.5~\mu$m which is about the measured radius of neutrophils, a main
type of leukocytes undergoing rolling adhesion \cite{hochmuth:93}.
\begin{table}[t]
    \begin{center}
    \begin{tabular}{lccc}
      \hline\hline
      Parameter $\rightarrow$& typical value& meaning&reference \\
      non-dimensionalized&(dimensionless)& &\\
      \hline
      $R \rightarrow 1$&$4.5\ldots5~\mu$m& radius&\cite{hochmuth:93,alon:97}\\
      $\dot\gamma \rightarrow 1$&$50\ldots150$~Hz&shear rate&\cite{springer:01}\\
      $T_a$&$293\ldots310$~K&ambient temperature&\\
      $\eta$&$1\ldots3$~Pa\,s&viscosity&\cite{springer:01}\\
      $\Delta \rho$&$50$~kg/m$^3$&density difference&\cite{munn:94}\\
      $\kk$&$10^{-5}\ldots10^{-2}$~N/m
      &bond spring constant&\cite{fritz:98,shao:98,hammer:92}\\
      $ \rightarrow \kk/6\pi\eta R \dot\gamma$&$(10^{-1}\ldots10^3)$&&\\
      $k_{on} $&$10^3\ldots10^4$~Hz&on-rate&\cite{schwarz:04}\\
      $\rightarrow \pi = k_{on}/\dot\gamma$&($10^{-3}$\ldots 10)&&\\
      $k_{0} $&$0.5\ldots 300$~Hz&unstressed off-rate&\cite{alon:97,schwarz:03}\\
      $\rightarrow \epsilon_0 = k_{0}/\dot\gamma$&($10^{-4}\ldots 10^3)$&&\\
      $r_0 $&$50$~nm&capture radius&\\
      $\rightarrow r_0/R$&($10^{-2}$)&&\\
      $d $&$0.1\ldots1~\mu$m&ligand-ligand distance&\cite{alon:95,alon:97}\\
      $ \rightarrow  d/R$&($0.02\ldots0.2$)&&\\
      $\rc $&$2\ldots4\cdot 10^{-11}$~m&reactive compliance&\cite{alon:97}\\
      $\rightarrow 6\pi\eta R^2 \dot\gamma \rc/k_BT_a$&(0.1\ldots0.6)&&\\
      $N_r$&$50\ldots5000$&Number of receptors&\cite{knutton:75,hammer:92,springer:99}\\
      $h_{min}$&$15$~nm&minimum cell height &\cite{bell:84}\\
      $\rightarrow h_{min}/R$&$(3\cdot 10^{-3})$&&\\
      \hline\hline
    \end{tabular}
    \end{center}
    \caption{Parameters used for the adhesive dynamics simulations. If
    no extra symbol for the dimensionless quantity is defined we use
    the same symbol for both the dimensional and dimensionless
    representation of this quantity.
      \label{tab:leuk:parameters}}
\end{table}
The $N_r$ receptor patches are randomly distributed on the cell surface. The
receptor patches might be identified with cell microvilli, which are membrane
projections to whose tips the selectin receptors are localized \cite{mcever:95}. The number of
microvilli on a leukocyte varies from several hundreds \cite{springer:99} up
to 10,000 \cite{knutton:75,hammer:92}.  In our simulations, we use $N_r$ in
the order of $10^3$. Although usually several receptors can be found on the
microvilli tips, we allow only one bond per receptor patch, i.\,e., $N_r$ is
also the total number of receptors. Receptors and ligands are spatially
extended in the nm-range and their binding sites diffuse within some region
around their linkage.  The diffusion constant of a nm-sized object is about
three orders of magnitude larger than that of the cell itself. Therefore, it is
sufficient to account for the spatial distribution of the location of binding
by introducing a capture sphere with capture radius $r_0 = 50$~nm, which is
about the combined length of ligand and receptor \cite{mcever:95a,bell:84}. In
regard to the ligands, we consider them being distributed on a square lattice
with lattice constant $d$. Here, $d$ is obtained from $d = \sqrt{1/N_l}$, with
$N_l$ being the average number of ligands per $\mu$m$^2$.  In flow chamber
experiments $N_l$ typically varies between $(1-100)/\mu$m$^2$
\cite{alon:95,alon:97}.

The on-rate $k_{on}$ for single bond formation from a receptor-ligand
encounter in \tab{tab:leuk:parameters} is given in units of Hz.
Experimentally, it is very difficult to determine this rate directly.  Using
three-dimensional affinity data, an upper limit has been estimated to be
$10^4$~Hz \cite{schwarz:04}. For the force dependent off-rate $k_{off}$ we
have the two Bell parameters $k_0, F_d$, where the detachment force is $F_d =
k_BT_a/\rc$ with reactive compliance $\rc$. Both, unstressed off-rate $k_0$
and reactive compliance $\rc$ have been measured for different selectin bonds
\cite{alon:95,alon:97}.  The reactive compliance for L-selectin bonds is $\rc
= 2\cdot 10^{-11}$~m \cite{alon:97}. This corresponds to a typical detachment
force of 200~pN.

For closed bonds we use the linear force extension curve explained in
\sec{sec:algo:addyn} with spring constant $\kk$. For the
P-selectin-\nolinebreak{PSGL-1} complex Fritz et al. \cite{fritz:98} measured
a value of $\kk_{RL} = 5.3 \cdot 10^{-3}$~N/m. Recently, also the role of
microvilli elasticity was discussed \cite{hammer:05}.  Shao et al.\ determined
the spring constant of microvilli in the low force regime to be $\kk_{mv} =
4.3 \cdot 10^{-5}$~N/m \cite{shao:98}.  The total spring constant $\kk =
\kk_{mv}\kk_{RL}/(\kk_{mv}+\kk_{RL})$ of the series of microvilli and bond would then
be dominated by the microvilli spring constant. Besides the bond forces, we
include only the buoyant force due to the small density difference $\Delta
\rho$ between the cell and the surrounding medium. Other nonspecific repulsion
forces arising from electrostatic and steric stabilization forces are
effectively taken into account by introducing a simulation rule that the cell
can approach the wall only up to a distance of $h_{min} = 15$~nm
\cite{bell:84}. In a physiological system, this would correspond to the
thickness of the glycocalyx, a protective layer of sugar covering the
endothelium in blood vessels.

\section{Definition of the dynamic states of motion}
\label{sec:def_states}

\subsection{Experiment and simulation}
\label{sec:exp_sim}

\begin{figure}
  \begin{center}
    \begin{tabular}{c@{\hspace{.04\linewidth}}c}
      \resizebox{.46\linewidth}{!}{\includegraphics{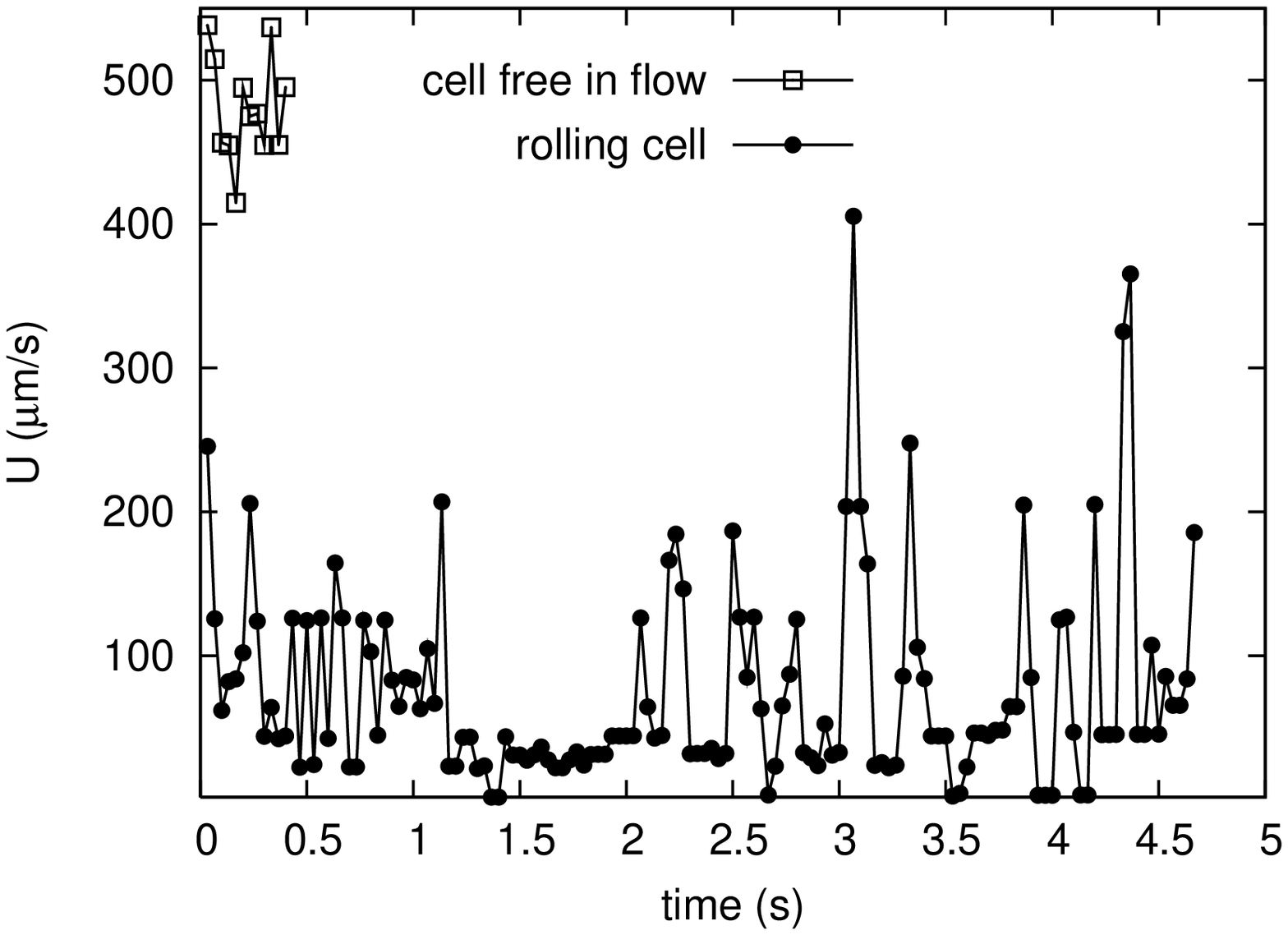}}&
      \resizebox{.46\linewidth}{!}{\includegraphics{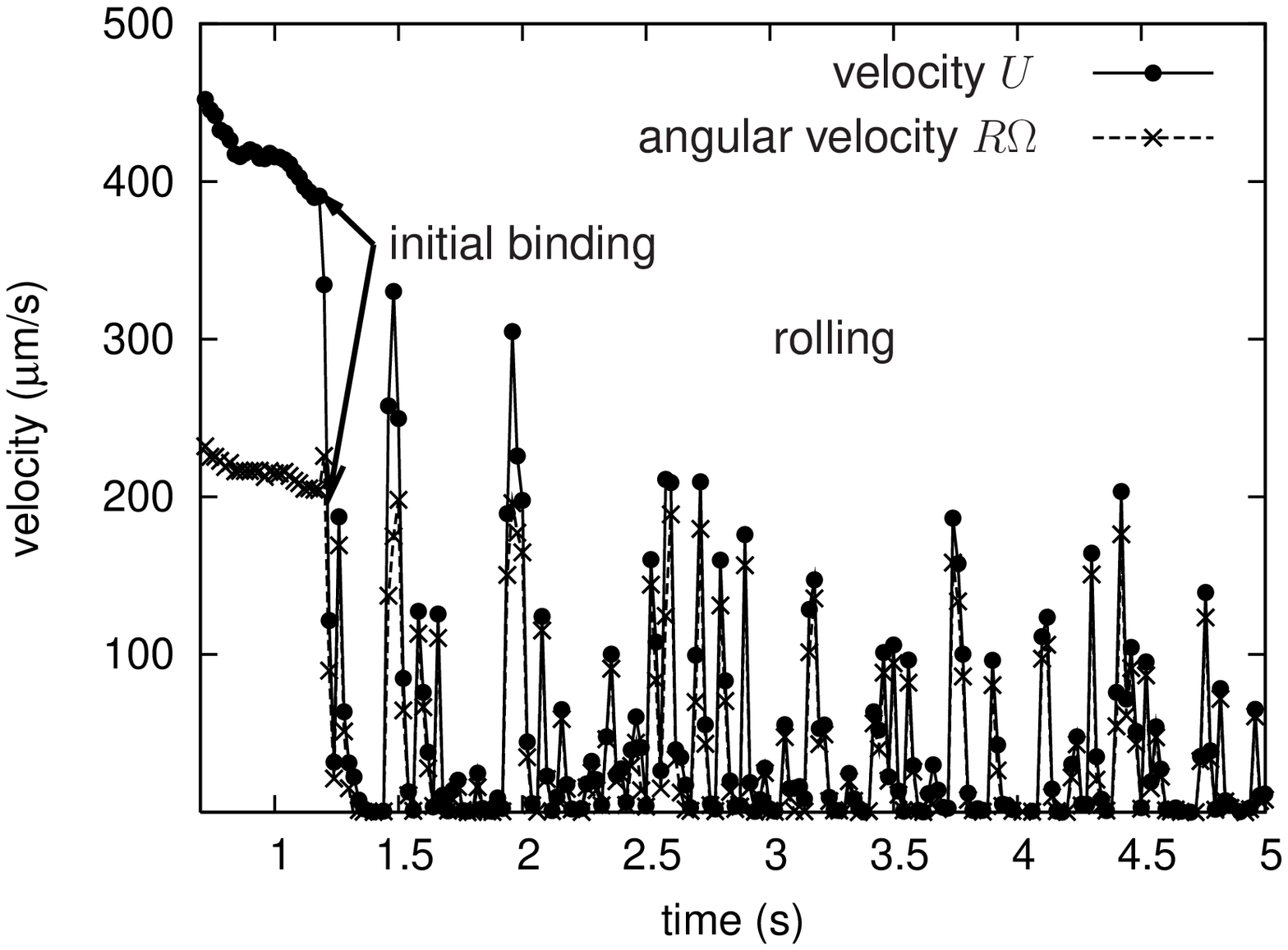}}\\
      (a) & (b) 
    \end{tabular}
    \caption{Snapshots of leukocyte motion. (a) Velocity of a
      leukocyte (neutrophil) rolling on L-selectin ligand PNAd with 60
      sites per $\mu$m$^2$. (Data obtained from Ref.~\cite[Fig.~1A]{alon:97}).  (b)
      Translational and angular velocity ($U$ and $R\Omega$,
      respectively) of a sphere with radius $R = 5~\mu$m measured in adhesive
      dynamics simulation at the same ligand density (with on-rate 
      $k_{on} = 60$~Hz and unstressed off-rate $k_{0} = 6.8$~Hz).
     \label{fig:leuk:expsnap}
    }
\end{center}
\end{figure}
Having completed the model definition, we now start to analyze the simulation
results. For the following we first have to clarify what is meant by the
velocity of the cell. As we include Brownian motion, the velocity of the cell
$U(t)$ cannot be its instantaneous velocity, because for the trajectory of a
Brownian particle $\lim_{\delta t \rightarrow 0}(X(t + \delta t) -
X(t))/\delta t$ is not a well-defined quantity \cite{horsthemke:84}. Thus, we
define the velocity of the sphere at time $t$ by a difference quotient $U(t)
:= (X(t + \Delta t) - X(t))/\Delta t$ with some time interval $\Delta t$.
Throughout this paper, we choose $\Delta t = 0.02$~s, which corresponds to a
camera resolution of $50$~Hz. The angular velocity for rotations about the
$y$-axis $\Omega(t)$ is defined in a similar way.

In \fig{fig:leuk:expsnap} we compare in a qualitative way the data
obtained in a numerical simulation of leukocyte motion with data of a
rolling cell obtained in a flow chamber experiment. In
\fig{fig:leuk:expsnap}a the translational velocity $U$ (denoting the
velocity in the direction of imposed shear flow) of a rolling
leukocyte for some period of time is shown as experimentally measured
by Alon and co-workers \cite{alon:97}. The rolling state is identified
by a strong decrease of the average cell velocity compared to the
average velocity of a cell moving freely in hydrodynamic flow.
\fig{fig:leuk:expsnap}b shows a representative trajectory from our
simulations. No attempt has been made to fit this
data set to the experimental one. Yet it is clear that both data sets
show very similar features, including the strong decrease of velocity
upon binding and the subsequent bursts in velocity, which correspond
to stochastic rupture events at the trailing edge allowing for cell
movement.  In contrast to the experimental data, the simulation data
also records the angular velocity $R \Omega$.  The simulation reveals
that the $R \Omega$ and $U$ curves collapse onto one curve as soon as
the cell binds for the first time (at $t \approx 1.2$ s). Before
initial binding, the cell slips over the substrate with $R\Omega/U <
0.57$ as discussed in \sec{sec:algo:stokes}. This observation
motivates us to define the dynamic state of \emph{rolling} by
$R\Omega/U \rightarrow 1$, in accordance with the common understanding
of this term in macroscopic mechanics. However, before we develop this
idea in more detail in \sec{sec:leuk:statedef}, we first show with an
analytical calculation how the action of a bond synchronizes
translational and rotational velocities.

\subsection{The stopping-process: emergence of rolling}
\label{sec:stop}

In order to understand how the cell comes to a stop after the first bond has
been formed, in the following we consider a set of simplified equations of
motion.
\begin{figure}[t!]
  \begin{center}
    \includegraphics[width=.66\linewidth]{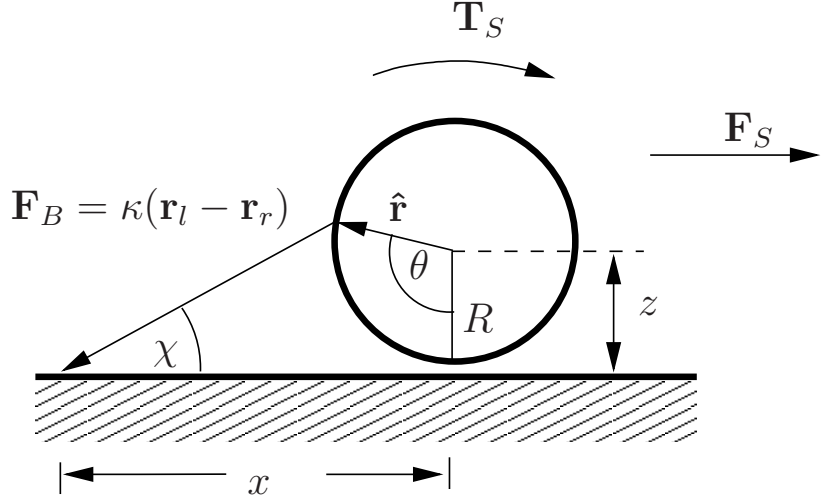}
    \caption{A cell is stopped by a tether force
      $\boldv{F}_B$. $\boldv{r}_l$ and $\boldv{r}_r$ are ligand and
      receptors position, respectively. $x, \theta$ define two
      degrees of freedom, $\chi$ the angle between the bond and the
      wall.  In equilibrium (i.\,e.\ $U = \dot x = 0, \Omega = \dot\theta = 0$) the
      shear force $\boldv{F}_S$ and the shear torque $\boldv{T}_S$
      are balanced by the tether force and its respective torque.
      Not shown are gravitational force and repulsive forces from
      the substrate that compensate all downward acting forces.}
    \label{fig:leuk:stopp}
  \end{center}
\end{figure}
We reduce our analysis to a two dimensional case in which the cell is only
allowed to translate in the x-direction and to rotate about the y-axis. These
two degrees of freedom are called $x$ for the translational and $\theta$ for
the rotational degree of freedom, respectively. In regard to the $z$-direction
we assume that the cell moves at constant height above the wall ($\dot z = 0$)
with $z = R + r_0$. Furthermore, we neglect Brownian motion, i.\,e., we
consider the limit $Pe \rightarrow \infty$.  Then, the equations of motion for
the two coordinates read (with $\dot x = U$ and $\dot\theta =
\Omega$, cf. \sec{sec:algo:stokes})
\begin{align}
  \label{eq:leuk:hdmotion}
  \left(\begin{array}{c}\dot x\\ \dot \theta \end{array}\right)
  = \mathsf{M}_{x\theta}
  \left(\begin{array}{c}  F_{B,x}\\ T_{B,y} \end{array}\right)
  +  \left(\begin{array}{c}  U_{hd}\\ \Omega_{hd} \end{array} \right),
\end{align}
with $F_{B,x}$ the $x$-component of the bond force and $T_{B,y}$ the
$y$-component of the torque which is due to the bond (see
\fig{fig:leuk:stopp}). Because the bond is modeled as a linear spring, it has
mechanical properties and we call it a \emph{tether}.  The matrix
$\mathsf{M}_{x\theta}$ denotes the $(x\theta)$-sector of the mobility matrix
introduced in \eq{langevin-ito}.  If no bond is formed, then $F_{B,x} =
T_{B,y} = 0$, and $U = U_{hd}$ and $\Omega = \Omega_{hd}$. The free velocity
$U_{hd}$ is often referred to as the \emph{hydrodynamic velocity} of the cell.
From \fig{fig:leuk:stopp} we read off tether force and tether torque
$\boldv{F}_B$ and $\boldv{T}_B$, respectively:
\begin{align}
  \label{eq:leuk:FTdef}
  \boldv{F}_B(t) = \kk\left(\!\begin{array}{c}
    -x(t) + \R\sin(\theta(t))\\ 0\\
    \R\cos(\theta(t)) - z
    \end{array}\!\right), \;\;
  \boldv{T}_B(t) =  \mathbf{\hat r} \times  \boldv{F}_B, \;\;
  \mathbf{\hat r} = 
  \left(\!\begin{array}{c}
    -\R\sin(\theta(t))\\
    0\\ 
    -\R\cos(\theta(t))
  \end{array}\!\right). 
\end{align}
In the following we distinguish two different cases with respect to the
resting length of the bonds. We start with the case were we assume the resting
length to be zero. With the cell moving at a constant height $z=R + r_0$, at
$t=0$ already a small bond force in $z$-direction exists (but no torque).
Briefly after bond formation the cell has not moved significantly and
$x/\Rr,\theta \ll 1$ holds.  Thus, the quantities $F_{B,x}, T_{B,y}$ can be
approximated as
\begin{align}
  \label{eq:leuk:FTapprox}
  F_{B,x} = \kk(\R\theta - x) + \mathcal{O}(\theta^3),\ha
  T_{B,y} = \kk(\R x - z \R\theta) + \mathcal{O}(\theta^3) + \mathcal{O}(x\theta^2).
\end{align} 
Reinserting these approximate expressions into \eq{eq:leuk:hdmotion}, we
obtain a first order linear differential equation for $x,\theta$
\begin{align}
  \label{eq:leuk:hdlinmotion}
  \left(\begin{array}{c}\dot x\\ \dot \theta \end{array}\right) = 
  \mathsf{C}
  \left(\begin{array}{c} x\\ \theta \end{array}\right) +
  \left(\begin{array}{c}  
    U_{hd}\\ \Omega_{hd} \end{array} \right),\ha
  \mathsf{C} := \mathsf{M}_{x\theta}\mathsf{X},\ha
  \mathsf{X} := \kk
 \left(\begin{array}{cc}
   -1 & \Rr\\  
    \Rr & -z\Rr 
 \end{array}\right),
\end{align}
which can readily be solved with the proper boundary conditions ($x(0) =
\theta(0) = 0$). The determinant of the matrix $\mathsf{C}$ is $\det\mathsf{C}
= \kappa^2R(z-R)\det{M}_{x\theta}$. From the dissipative nature of the
mobility matrix expressed, e.\,g., in \eq{noise}, it follows that
$\mathsf{M}_{x\theta}$ is positive definite, which is then also true for the
matrix $\mathsf{C}$. As the diagonal elements of $\mathsf{M}_{x\theta}$ are
significantly larger than the off-diagonal element \cite{jones:92,jones:98},
one easily finds that $\tr\mathsf{C}$ is negative. Therefore, $\mathsf{C}$ has
two negative eigenvalues, which we denote in the following by
$\lambda_\pm$. As $z-R\ll R$ one can approximate $\lambda_+ \approx
\tr\mathsf{C}$ and $\lambda_- \approx \det\mathsf{C}/\tr\mathsf{C}$.  These
two eigenvalues represent two different timescales with $|\lambda_-| \ll
|\lambda_+|$. With this the solution to \eq{eq:leuk:hdlinmotion} can be
written as
\begin{align}
  \label{eq:leuk:xthsol02}
   \left(\!\!\begin{array}{c}x(t)\\\theta(t)\end{array}\!\!\right) =
  \left(\!\!\begin{array}{c} 
    1\\     \frac{\lambda_+ - (\mathsf{C})_{11}}{(\mathsf{C})_{12}}
  \end{array}\!\!\right)\frac{U_{hd} + \lambda_-x_\infty}{\lambda_- - \lambda_+} 
    \left(1 - e^{\lambda_{+} t}\right) -      
  \left(\!\!\begin{array}{c} 
    1\\     \frac{\lambda_{-} -  (\mathsf{C})_{11}}{ (\mathsf{C})_{12}}
  \end{array}\!\!\right)\frac{U_{hd} + \lambda_+ x_\infty}{\lambda_- - \lambda_+} 
    \left(1 - e^{\lambda_{-} t}\right).
\end{align}
where the asymptotic solution after the cell has stopped is given by
\begin{align}
  \label{eq:leuk:xthsol01}
  \left(\begin{array}{c} x_\infty\\  \theta_\infty \end{array}\right) 
  := -(\mathsf{C})^{-1}\left(\begin{array}{c}  U_{hd}\\ \Omega_{hd} \end{array} \right).
\end{align}
\eq{eq:leuk:xthsol01} is the linearized version of the force and torque
balance condition at mechanical equilibrium. In non-linearized form the force
and torque balance equation reads
\begin{align*}
   \left(\begin{array}{c} F_{B,x}\\  T_{B,y} \end{array}\right) 
  = -\mathsf{M}^{-1}_{x\theta}\left(\begin{array}{c}  U_{hd}\\ \Omega_{hd} \end{array} \right). 
\end{align*}
Then, for $z = 1.01R$ one gets $F_{B,x} \approx -1.7\cdot 6\pi\eta\dot\gamma R^2$
and $T_{B,y} \approx -0.6 \cdot 6\pi\eta\dot\gamma R^3$ \cite{goldman:67b}.
The relation to the quantities $x_\infty,\chi$ is given in Ref.~\cite{alon:97} 
\begin{align}
  \label{eq:ftbalance2}
  \|\boldv{F}_B\|\cos\chi = \|\boldv{F}_S\|, \ha \|\boldv{F}_B\|
  x_\infty \sin\chi = \|\boldv{T}_S \| + R \|\boldv{F}_S \|.
\end{align}
The relation with $\theta_\infty$ is given by the purely geometrical
relation $x_\infty/R = \sin\theta_\infty + (1 - \cos\theta_\infty)/\tan\chi$
(cf. \fig{fig:leuk:stopp}).  In contrast to the linear version,
\eq{eq:ftbalance2} cannot be solved for $x_\infty, \theta_\infty$.

\begin{figure}[t!]
  \begin{center}
    \begin{tabular}{c@{\hspace{.04\linewidth}}c}
      \resizebox{.46\linewidth}{!}{\includegraphics{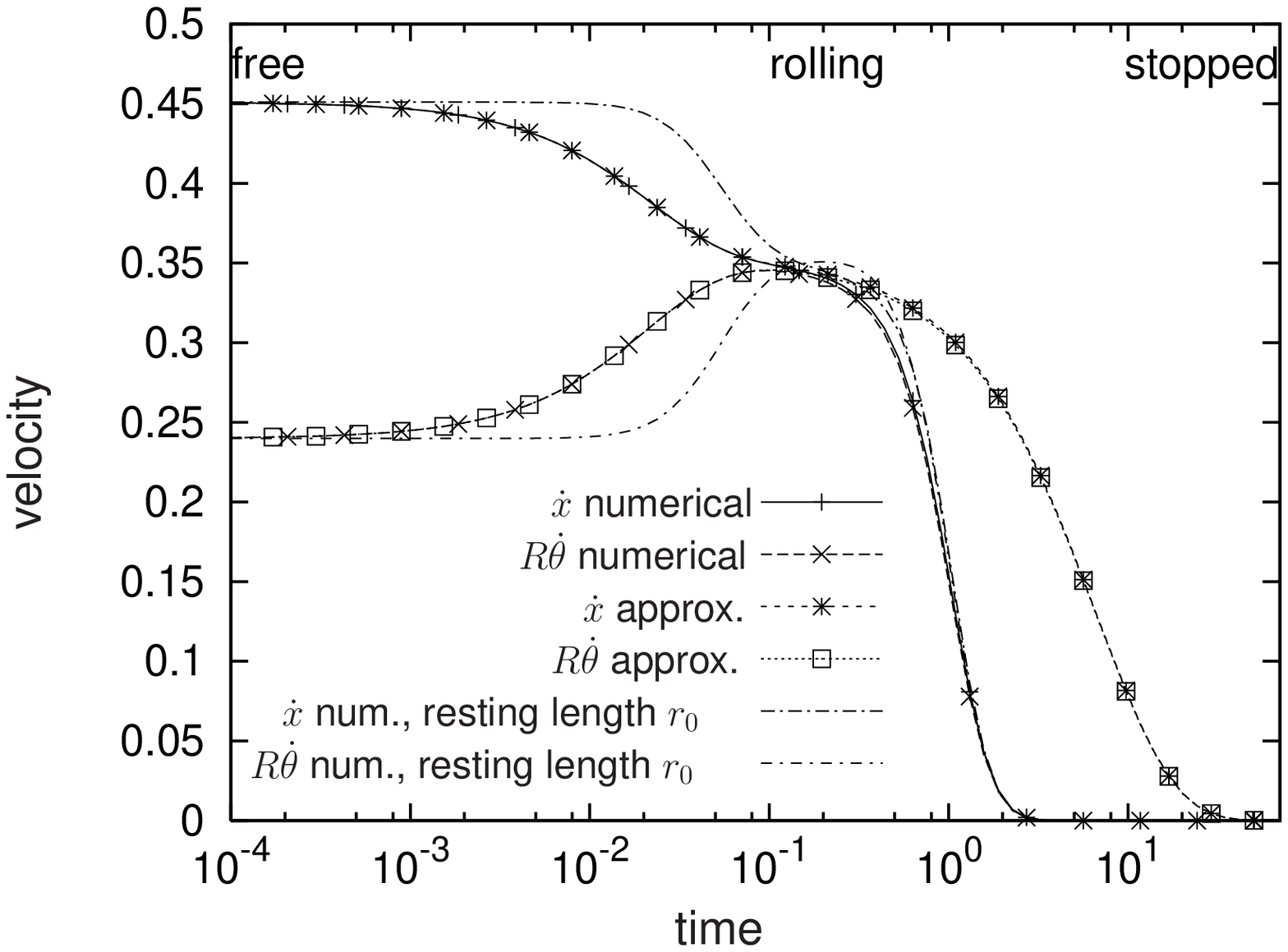}}&
      \resizebox{.46\linewidth}{!}{\includegraphics{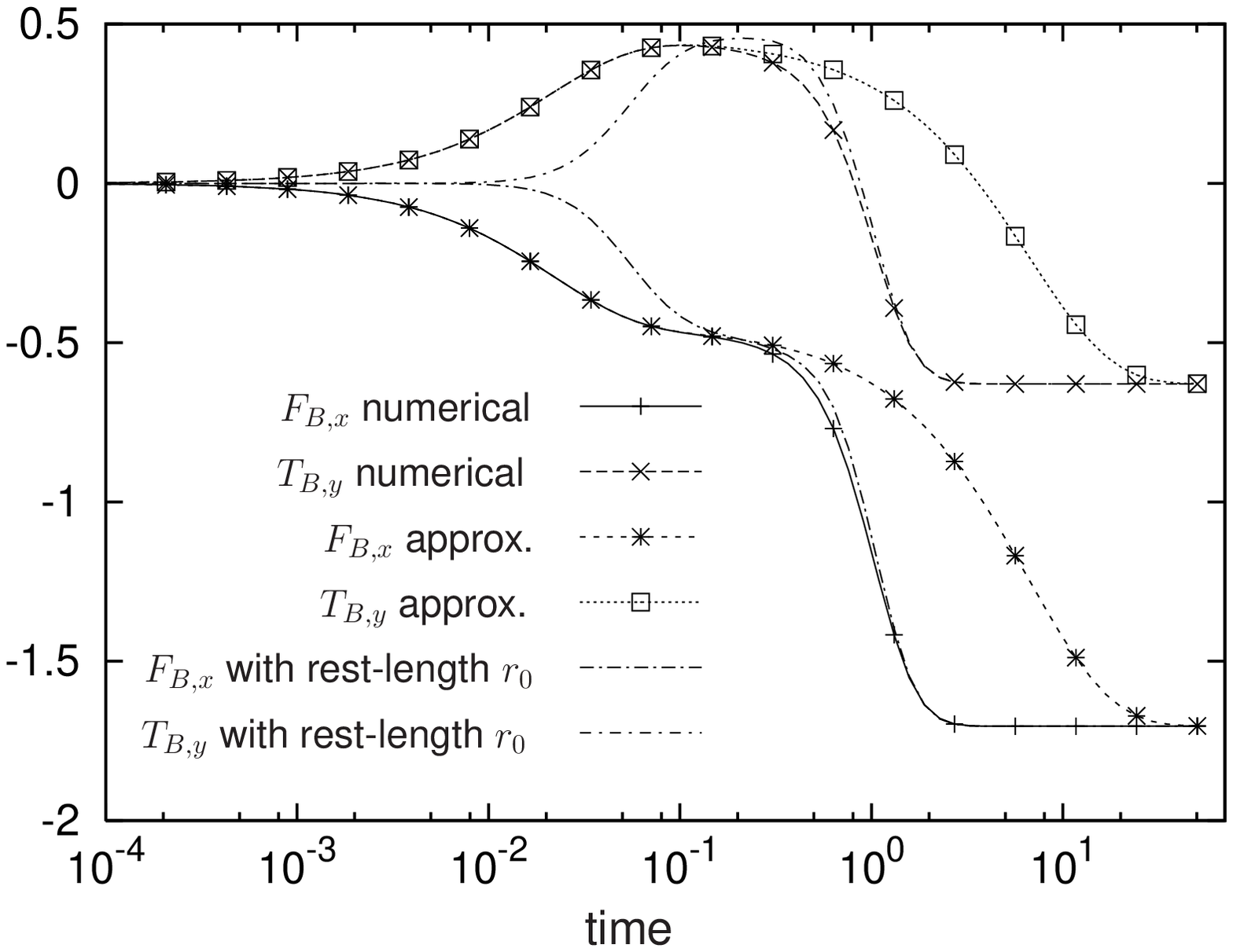}}\\
      (a) & (b) 
    \end{tabular}
    \caption{Comparison of the analytical solution
      for the stop dynamics mediated by a single bond
      \eq{eq:leuk:xthsol02} with values obtained from numerical
      integration of \eq{eq:leuk:hdmotion} ($z = 1.01\R$).  All
      plotted quantities are in dimensionless form, i.\,e., force is
      expressed in terms of $6\pi\eta R^2\dot\gamma$, torque in terms
      of $6\pi\eta R^3\dot\gamma$, velocity in $\dot\gamma R$, and
      time in $1/\dot\gamma$. For the dimensionless spring constant
      $\kk=10^{2}$ was used. For a set of typical parameter values: $R
      = 5~\mu$m, $\dot\gamma = 100$~Hz, $\eta = 10^{-3}$~Pa\,s, $\kk = 1$
      corresponds to $10^{-5}$~N/m.  (a) Plot of the (angular) velocity
      $\dot x$ ($\dot\theta$) as a function 
      of time for three different calculations: from
      numerical integration of \eq{eq:leuk:hdmotion}, from
      analytical derivative of \eq{eq:leuk:xthsol02}, and from
      numerical integration of \eq{eq:leuk:hdmotion} under the
      assumption that the initial bond length $r_0$ is the resting length of
      the bond $l_0 = r_0$. For the time axis a logarithmic scale is used.  (b)
      The same for the bond force (torque) $F_{B,x}$ ($T_{B,y}$).
     \label{fig:leuk:analyt01}
   }
\end{center}
\end{figure}

The initial velocities of the cell are the free hydrodynamic velocities with
$\R\dot\theta(0)/\dot x(0) \approx 0.5$ so that $x$ and $\theta$ do not
increase at the same speed (see \fig{fig:leuk:analyt01}a). Shortly after bond
formation, $t \ll 1/|\lambda_+|$, $\exp(\lambda_+ t) \approx 0$ and the time
development of $x,\theta$ is governed by the second term in
\eq{eq:leuk:xthsol02}. At intermediate times $t_{int}$ with $1/|\lambda_+| \ll
t_{int} \ll 1/|\lambda_-|$ the velocities are approximately given by
\begin{align}
    \left(\begin{array}{c}\dot x\\ \dot \theta \end{array}\right) \approx \lambda_- 
      \left(\!\begin{array}{c} 
    1\\     \frac{\lambda_{-} -  (\mathsf{C})_{11}}{ (\mathsf{C})_{12}}
  \end{array}\!\right)\frac{U_{hd} + \lambda_+ x_\infty}{\lambda_- - \lambda_+} 
      e^{\lambda_{-} t_{int}}.
\end{align}
Expanding $\R \dot\theta/\dot x$ in powers of $(z-\R)/\R\ll 1$, we obtain $\R
\dot\theta/\dot x = 1 + \mathcal{O}((z-\R)/\R)$ which reflects the definition
of rolling in mechanics. Thus, by the action of force and torque resulting
from the tether bond, translational and angular velocity of the cell are
adjusted and the cell starts rolling a short time after the first bond is
formed.  This can be seen in \fig{fig:leuk:analyt01}a, there the two
velocities $\dot x, \dot \theta$ are plotted as a function of time.
Although \eq{eq:leuk:xthsol02} correctly predicts the cell to arrest at large
times, the linear approximation \eq{eq:leuk:FTapprox} is not applicable at
these times, as can be seen from \fig{fig:leuk:analyt01}.  In
\fig{fig:leuk:analyt01}a the cell velocities as predicted form the analytical
solution to \eq{eq:leuk:hdlinmotion} are compared to the velocities which are
obtained when \eq{eq:leuk:hdmotion} is numerically integrated. We see that for
short times the linear approximation works quite well, until the cell has
reached its rolling state (i.\,e., $\R\Omega/U \approx 1$). At long time
scales the approximate solution provides only a qualitative prediction of the
stopping process. The figure shows that the higher order terms contributing to
the bond force/torque lead to a much faster stopping than predicted by the
approximation of these quantities with respect to first order in $x,\theta$
(see \eq{eq:leuk:FTapprox}).

We now briefly discuss the case of a non-vanishing resting length $l_0$ of the
bond (a non-vanishing resting length with $l_0 \leq r_0$ is used in
our simulations as explained in the Appendix). For this we assume the height
$z - \Rr$ of the cell above the wall to be equal to the resting length of the
tether bond. In that case the spring constant in \eq{eq:leuk:FTdef} is
replaced by $\kk \rightarrow \kk (1 - l_0/\|\boldv{r}_l-\boldv{r}_r\|)$ (see
\fig{fig:leuk:stopp}).  The leading order term in the power series of
$F_{B,x}$ is then of second order in $x,\theta$. Thus the linear approximation
of the equations of motion cannot be applied in this case and therefore we consider
this case only numerically. The results are also shown in
\fig{fig:leuk:analyt01}a. We note that the qualitative behavior is very
similar to the case of zero bond resting length. \fig{fig:leuk:analyt01}a
shows that at short times the cell moves on with almost unchanged velocities,
then the velocities adjust to $\R\dot\theta \approx \dot x$ (for some time
even $\R\dot\theta > \dot x$) and finally the cell stops after about the same
time as in the case with zero bond resting length.

The linearized analysis of the rather complex motion of the cell under the
action of a single bond might appear to be somehow simplifying, but it
nicely reveals the mechanism that leads to cell rolling.  This can best be
understood from the time dependence of the torque that is exerted by the
tether, see \fig{fig:leuk:analyt01}b.  In our approximation the torque is
given by $T_{B,y} = \kk (\R x - z\R\theta)$ (see \eq{eq:leuk:FTapprox}).
Initially, $x$ increases faster than $\R\theta$ and as $z \approx \Rr$ the
torque is positive, i.\,e., it supports the shear torque and the cell starts
turning faster. At the same time the force $\R F_{B,x} \approx -T_{B,y}$ slows
down the translational motion.  The maximum torque $\dot T_{B,y} = 0$ is
reached when $z \dot \theta = \dot x$, i.\,e., when the cell is approximately
rolling. From that time on $x$ and $\theta$ increase at approximately the same
speed and eventually the torque will become negative (if $x < z\theta$) and
will act against the shear torque.  Similar arguments hold when repeating the
previous discussion with the exact expression for the torque given in
\eq{eq:leuk:FTdef}.
 
\subsection{Cell motion at multiple bonds: classification of states of motion}
\label{sec:leuk:statedef}

\begin{figure}[t!]
  \begin{center}
    \vspace*{-.5cm}
    \begin{tabular}{c@{\hspace{-.04\linewidth}}c}
      \resizebox{.59\linewidth}{!}{\includegraphics{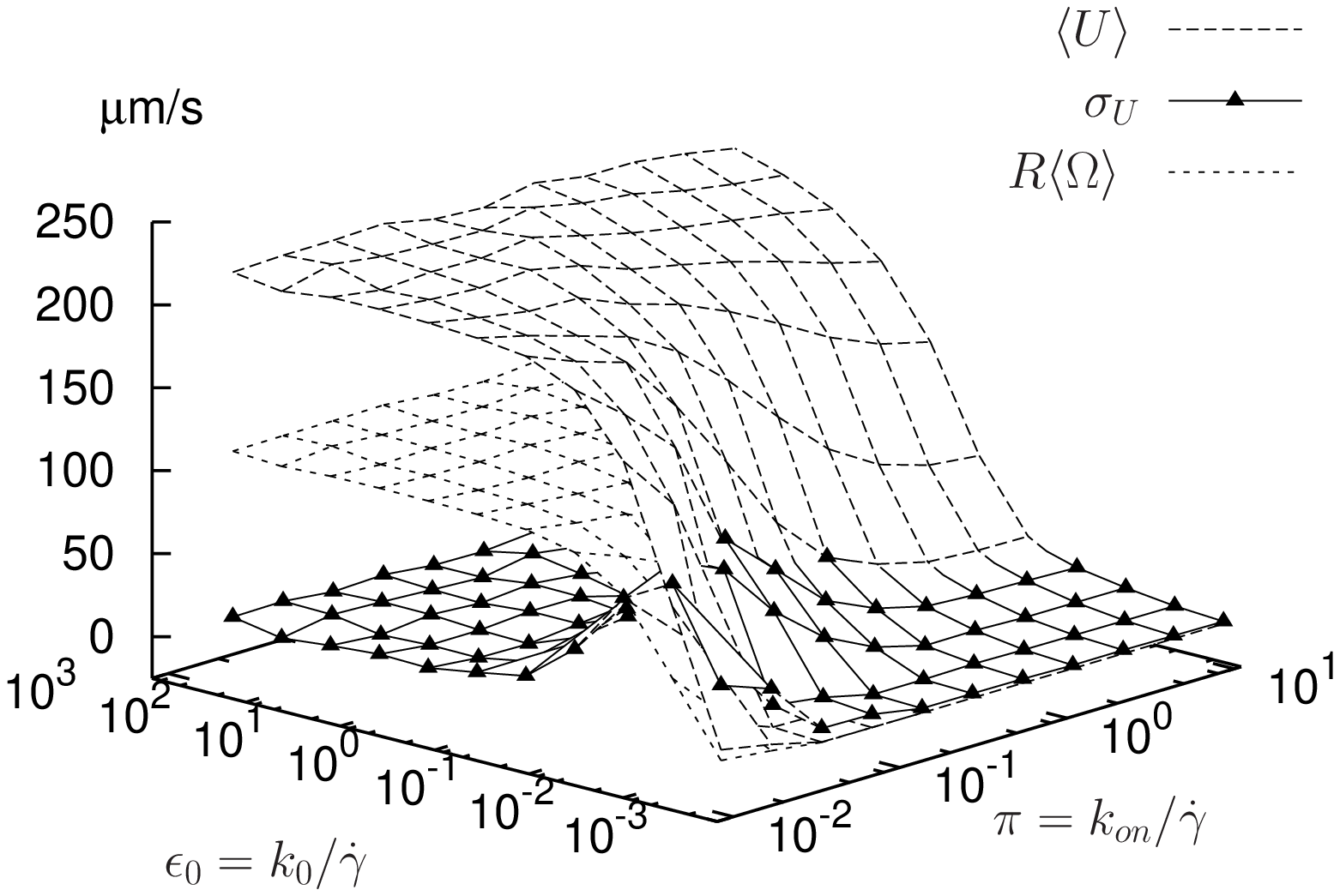}}&
      \raisebox{.7cm}[-.7cm]{%
      \resizebox{.44\linewidth}{!}{\includegraphics{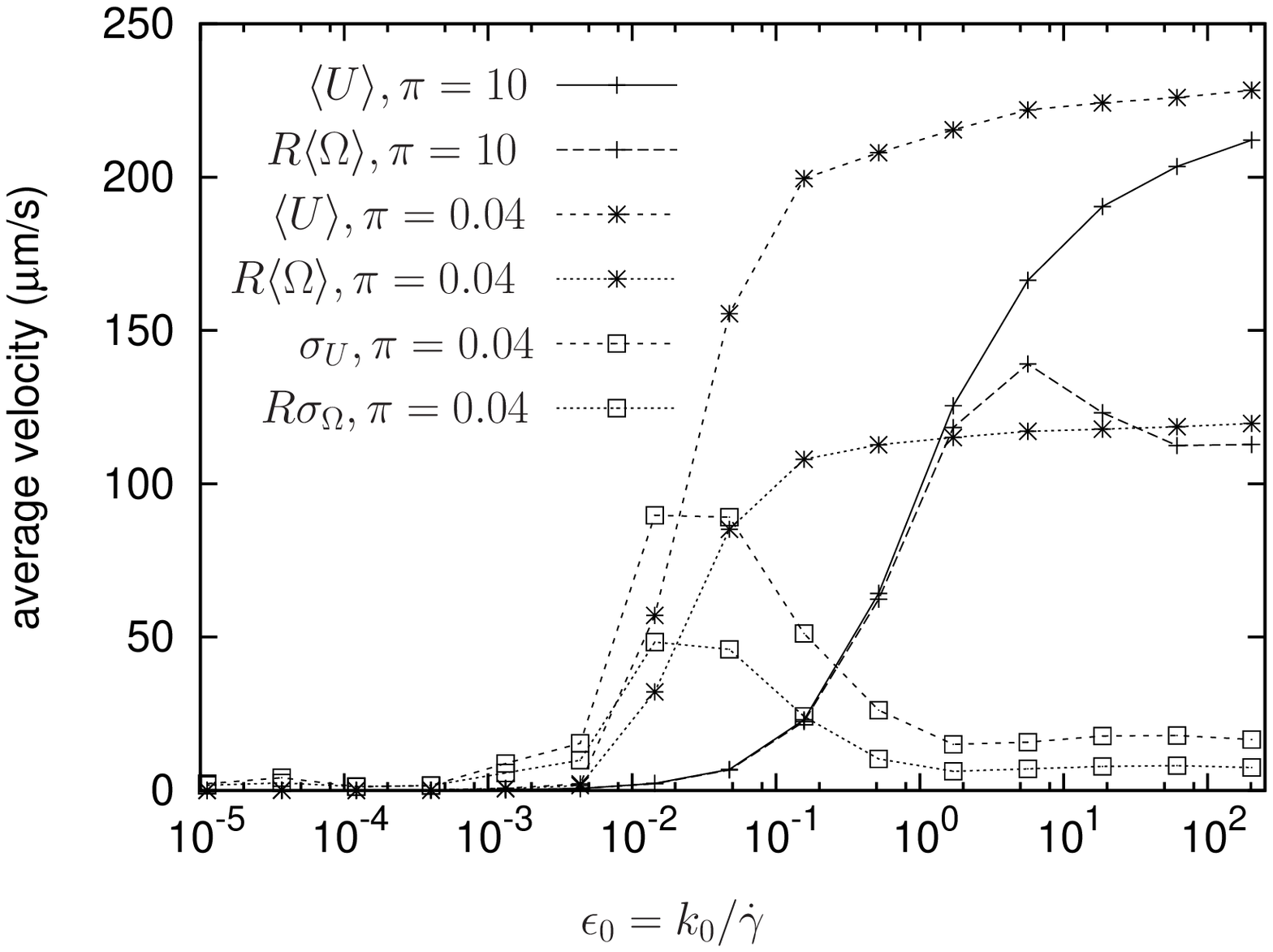}}}\\
      (a) & (b) \\
    \end{tabular}
    \caption{(a) Mean velocity $\mean{U}$, its
      standard deviation $\sigma_U$, and mean angular velocity
      $R\mean{\Omega}$ as functions of the dimensionless rates
      $\pi,\epsilon_0$. (b) $\mean{U}, R\mean{\Omega}, \sigma_U,
      \sigma_{R\Omega}$ as functions of the unstressed off-rate $\epsilon_0$
      for two different on-rates $\pi$. Parameters used for these simulations: $R =
      4.5\cdot 10^{-6}$~m, $T = 293$~K, $\dot\gamma = 100$~Hz, $\Delta\rho =
      0.05\cdot 10^{3}$~kg/m$^3$, $\eta = 1.002\cdot 10^{-3}$~Pa\,s, $\kk =
      1\cdot 10^{-3}$~N/m, $r_0= 1.0\cdot 10^{-2}R$ , $d = 5\cdot 10^{-2}R$,
      $N_r = 5000$, $\rc = 2\cdot 10^{-11}$~m.  The average was obtained over
      ten simulation runs of 20~s duration.
     \label{fig:leuk:3dvonoff}
    }
\end{center}
\end{figure}
The previous analysis did only include a single bond which in addition was not
allowed to rupture. In the presence of multiple bonds permanently forming and
rupturing the situation is much more complex.  Whether a cell is able to roll
or not depends then on the one hand on \emph{external} parameters like
ligand-density, shear rate and viscosity. On the other hand it depends also on
the \emph{internal} parameters of the single receptor-ligand complex, which in
our model are the on-rate $k_{on}$, the off-rate $k_{0}$ and the detachment
force $F_d$. In the following we will present our results mainly as a function
of the two internal rates in their dimensionless form, $\pi =
k_{on}/\dot\gamma$ and $\epsilon_0 = k_{0}/\dot\gamma$. In
\fig{fig:leuk:3dvonoff}a we plot the mean translational and angular velocities
as well as the standard deviation of the translational velocity, $\sigma_U =
\sqrt{\mean{U^2} - \mean{U}^2}$ (where the average is an average over time and
an ensemble of cells), in a large range of values for the dimensionless on-
and off-rates. To further illustrate the dependence of the kinetic quantities
on the on- and off-rate, in \fig{fig:leuk:3dvonoff}b the
$\epsilon_0$-dependence at fixed on-rate $\pi$ is re-plotted.  At $\pi = 10$,
one nicely sees that with decreasing $\epsilon_0$, translational- and angular
velocities first approximate each other, i.\,e., $U$ decreases and $R\Omega$
increases. Then, both together decrease to zero at very low off-rates.  At smaller
on-rates ($\pi = 0.04$), both $U$ and $R\Omega$ monotonically decrease with
decreasing off-rate and $R\Omega \approx U$ occurs only when both quantities
are close to zero.  The standard deviations of the velocities $\sigma_{U},
\sigma_{R\Omega}$ are small for very low and very high off-rates. In between,
they pass through a maximum which is located exactly at the transition from
unperturbed motion to cell arrest ($U \approx R\Omega \approx 0$).
\begin{table}[t]
    \begin{center}
    \begin{tabular}{l|cc}
      \hline \hline state&\multicolumn{2}{c}{definition}\\ 
      \hline\hline 
      free motion&\multicolumn{2}{c}{$\mean{U} > 0.95~U_{hd}$}\\ rolling
      adhesion&\multicolumn{2}{c}{$R\mean{\Omega}/\mean{U} > 0.8$  AND
      $0.95 > \mean{U}/U_{hd} > 0.01$}\\ firm
      adhesion&\multicolumn{2}{c}{$\mean{U} < 0.01~U_{hd}$}\\
      transient adhesion I&& AND $\sigma_U/\mean{U} < 0.5$\\ transient
      adhesion II&\raisebox{.2cm}[-.2cm]{$0.01<\mean{U}/U_{hd}$ AND
      $R\mean{\Omega}/\mean{U} < 0.8$}& AND $\sigma_U/\mean{U} >
      0.5$\\ \hline\hline
    \end{tabular}
    \end{center}
    \caption{Five stationary states of leukocyte motion.
      \label{tab:leuk:states}}
\end{table}
We now summarize these qualitative observations by defining five different
classes of stationary states of cell motion (see also \tab{tab:leuk:states}).

\begin{figure}[t!]
  \begin{center}
    \begin{tabular}{c@{\hspace{.04\linewidth}}c}
      \resizebox{.46\linewidth}{!}{\includegraphics{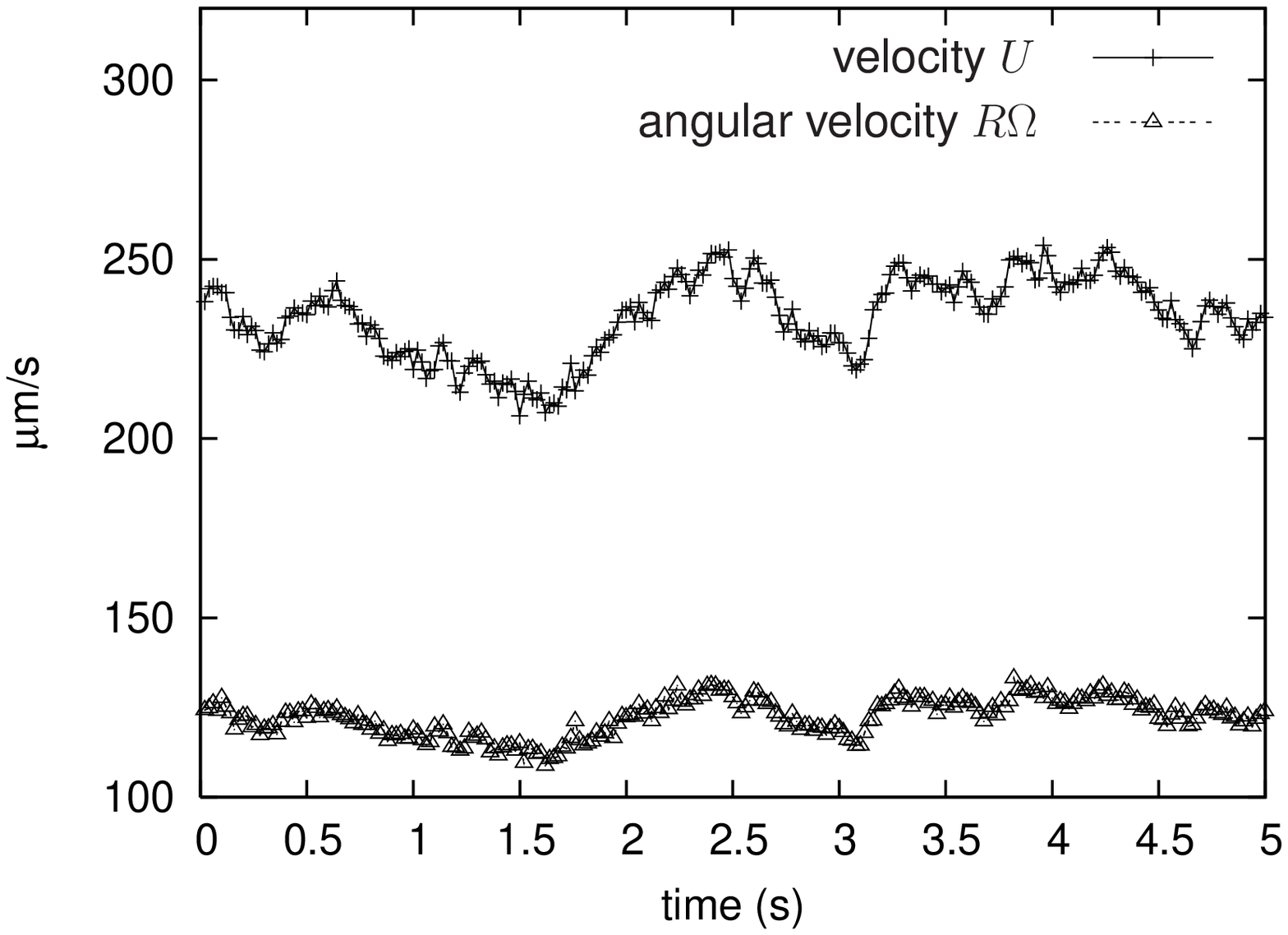}}&
      \resizebox{.46\linewidth}{!}{\includegraphics{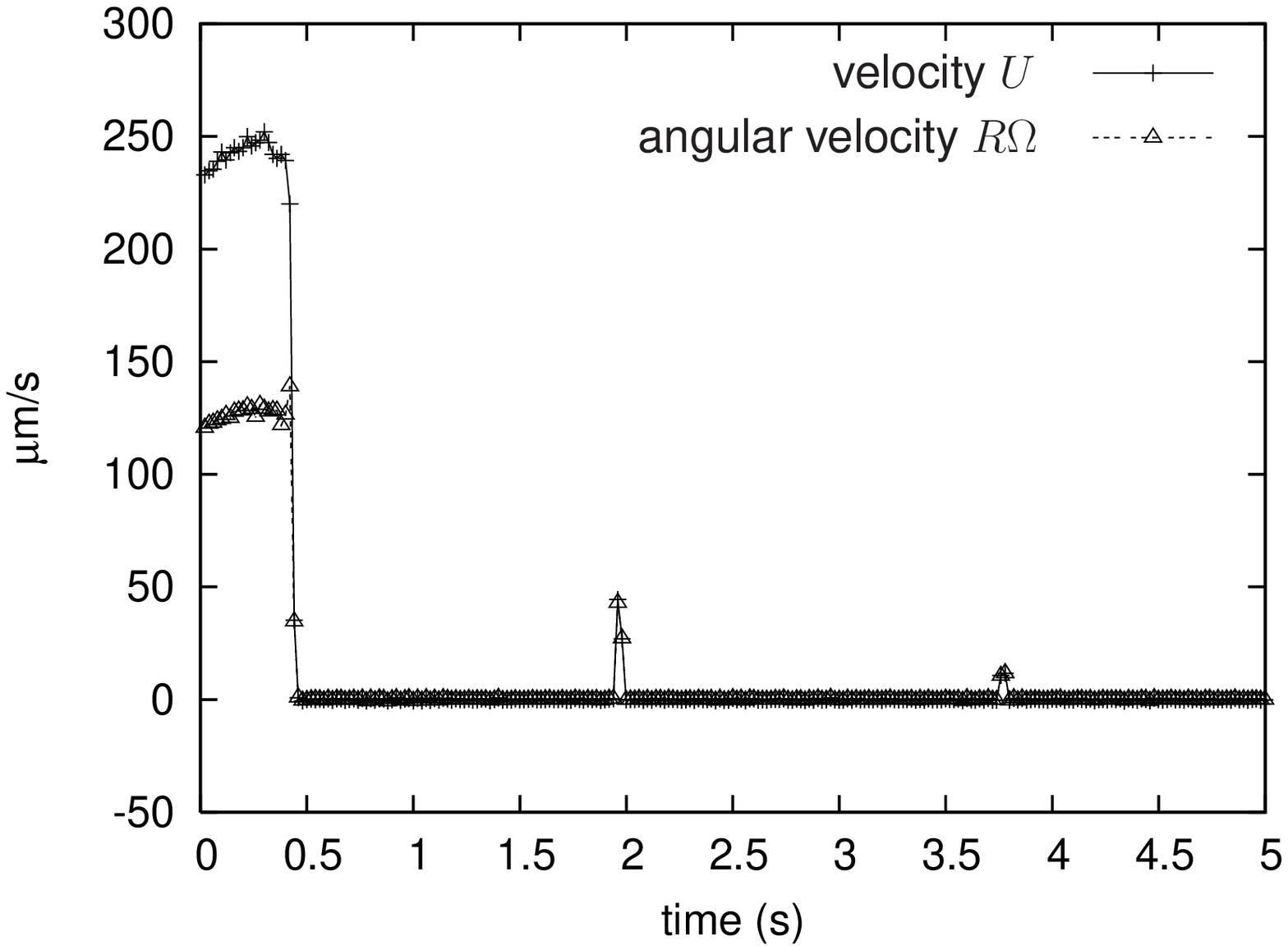}}\\
      (1) & (2) \\
      \resizebox{.46\linewidth}{!}{\includegraphics{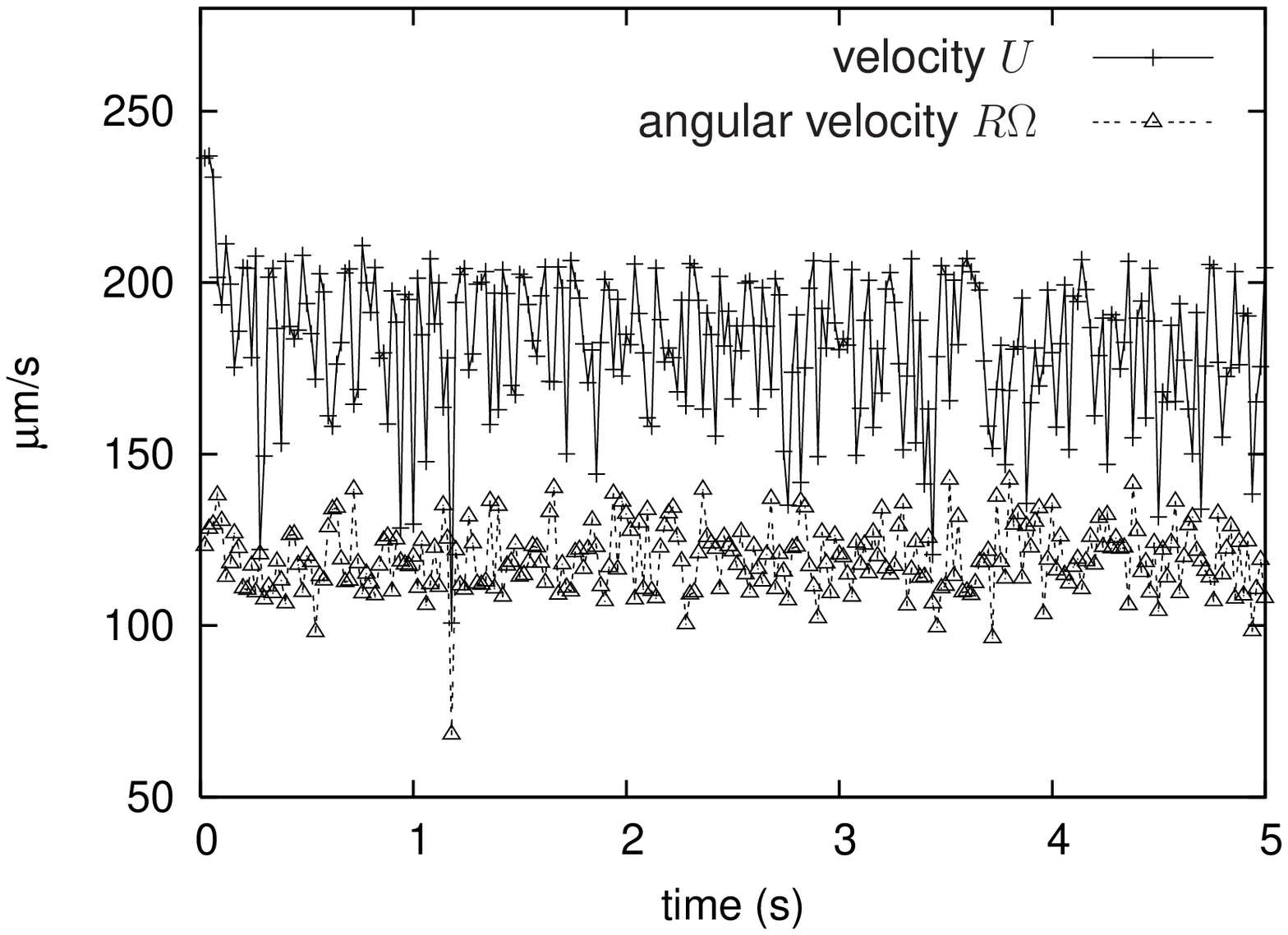}}&
      \resizebox{.46\linewidth}{!}{\includegraphics{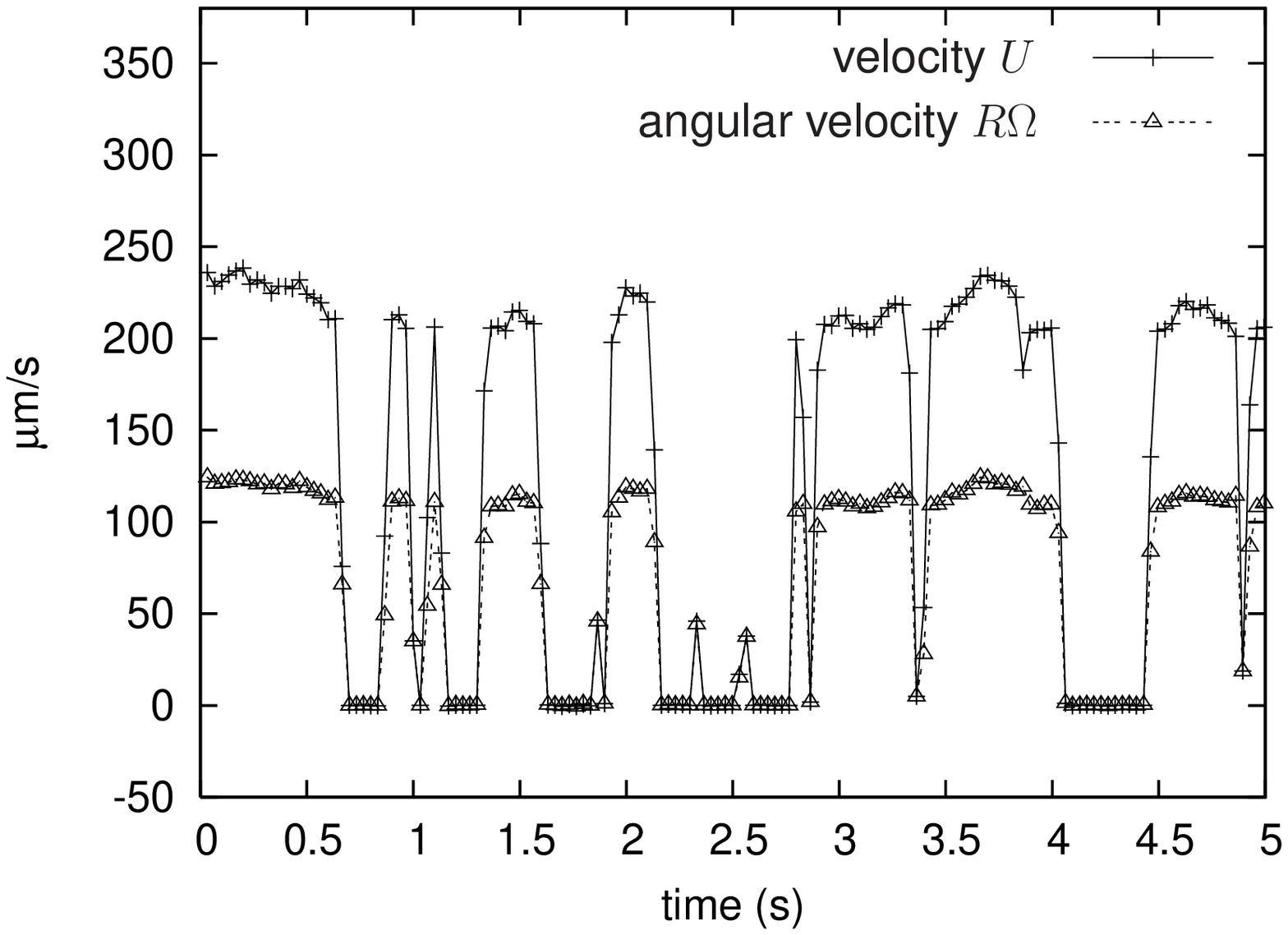}}\\
      (3) & (4) \\
      \resizebox{.46\linewidth}{!}{\includegraphics{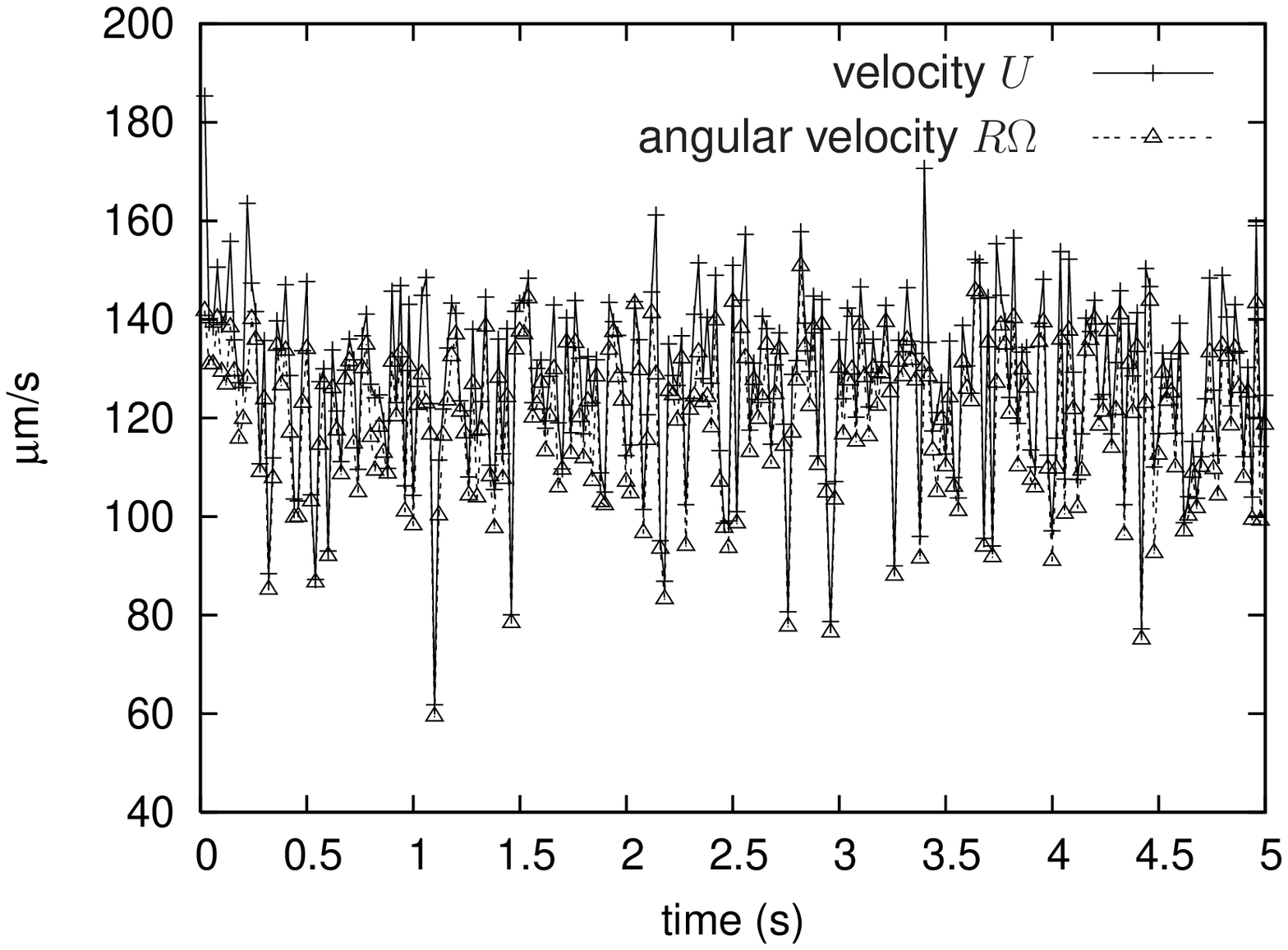}}&
      \resizebox{.46\linewidth}{!}{\includegraphics{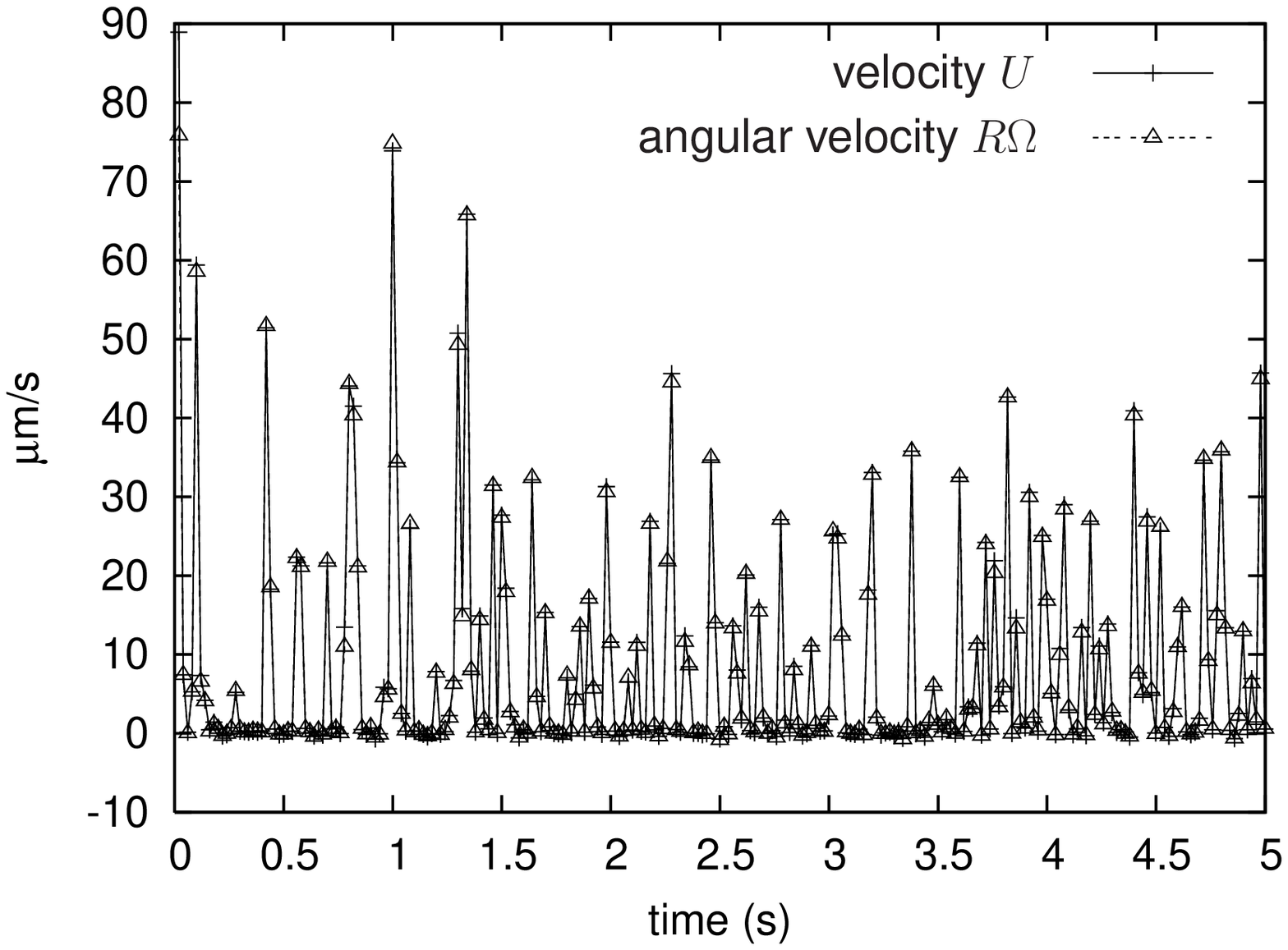}}\\
      (5) & (6) 
    \end{tabular}
    \caption{ Translational and angular velocities $U$ and $R\Omega$,
      respectively, that give examples of the different states of leukocyte
      motion defined in this section.  The labels (1-6) refer to the numbered
      points in the state diagram shown in \fig{fig:leuk:on-off-state}.  (1)
      Free motion, (2) firm adhesion, (3) `transient I', (4) `transient II',
      (5,6) rolling adhesion.
      \label{fig:leuk:phaseexamples}}
\end{center}
\end{figure}

\emph{Free motion}: We call a cell to move freely if its speed is larger than
$0.95~U_{hd}$ (an example is shown in \fig{fig:leuk:phaseexamples}(1)).  Free
motion as we define it does not imply that there are no bonds at all. The
definition given by us rather allows also for bonds with a very fast
dissociation rate (off-rate) or very small detachment forces. In this case
existing bonds dissociate before they are stretched enough to apply forces
that slow down the mean velocity of the cell below 95~\% of $U_{hd}$.
 
\emph{Firm adhesion} (arrest): This is the state when the mean translational
velocity $\mean{U}$ is less than $0.01 U_{hd}$
(\fig{fig:leuk:phaseexamples}(2)). This still allows for small jumps due to
rare dissociation events.  Besides that tether bonds compensate shear force
and torque (cf. \sec{sec:stop}).
 
\emph{Rolling adhesion}: The ratio $R\mean{\Omega}/\mean{U}$ is larger than
0.8.  As was shown in \sec{sec:algo:stokes}, this is well above the
hydrodynamic maximum of this ratio in the limit $z\rightarrow \R$ (i.\,e.,
when the cell touches the wall).  \fig{fig:leuk:phaseexamples}(5) and
\fig{fig:leuk:phaseexamples}(6) show two examples of computational leukocyte
rolling.

\emph{Transient adhesion}: If none of these criteria applies we define the
state as being transient.  Within this category we distinguish two sub-classes
according to the standard deviation $\sigma_U$.  By $\sigma_U/\mean{U} < 0.5$
the first sub-class (\emph{transient I}) is defined, otherwise the cell's
motion is in the sub-class \emph{transient II}.  `Transient I' occurs if bonds
form and rupture permanently, so that they reduce the (translational) velocity
considerably below the hydrodynamic velocity. However, in this case the bonds
do not last sufficiently long as to increase the ratio $R\Omega/U$ above
0.8. \fig{fig:leuk:phaseexamples}(3) shows an example for this kind of
motion. `Transient II' is characterized by alternating periods of arrest and
free motion which is illustrated in \fig{fig:leuk:phaseexamples}(4).

As the kinetic quantities $\mean{U},\mean{\Omega},\sigma_U$ vary
continously with respect to $\pi$ and $\epsilon_0$, the classification
given above is not unique. But it allows to clearly distinguish these
states in an \emph{on-off state diagram}, i.\,e., the states are in
general not degenerated.  Other classifications of leukocyte states
have been given before. For example, in the first paper on adhesive
dynamics also five states of motion were defined \cite{hammer:92}.  In
contrast to our definition, however, this classification was only
qualitative. In computer simulations of adhesive dynamics or other
numerical models for rolling adhesion, it is common to define states
of motion like \emph{free}, \emph{rolling} and \emph{firm adhesion},
but usually this is done like in an experimental context, which means
that these classifications are based only on the mean translational
velocity \cite{hammer:04a,hammer:05}. A classification into
\emph{immobile}, \emph{rolling} and \emph{detached} and a
corresponding state diagram has also been given by Bruinsma in a mean
field approach, which did not model in detail how force is distributed
over the molecular bonds \cite{bruinsma:96}.

\section{State diagram of leukocyte motion}
\label{sec:generalstate}

In order to determine the stationary state of motion for a given set of
parameters, we repeatedly performed the following computer simulation.
At a given set of parameter values we let the cell start at a height $z = \Rr
+ r_0$ and subsequently simulate its motion for 20~s.  As a result of the
downward acting buoyant force, which drives the cell even closer to the wall,
the wall ligands will be immediately within the capture range of the
cell-receptors. Then, as shown in \cite{korn:06}, the mean time for
receptor-ligand encounter is close to zero for typical ligand and receptor
densities found for leukocytes.  Therefore, cell-wall interactions arising
from bonds are assumed to influence the cell motion for the complete run of
the simulation.  To nevertheless rule out any initialization effects the mean
values and variances for $U,\Omega$ are only calculated for times greater than
$4$~s, i.\,e. 20\% of the total run length of $20$~s. To ensure proper
classification of the state of motion, each simulation run is repeated at
least ten times (each time with another randomly chosen receptor distribution)
and each run contributes to the mean values $\mean{U},\mean{\Omega}$ and their
standard deviations $\sigma_U, \sigma_\Omega$ (in the case that $\sigma_U$ is
large even more than ten simulation runs were performed). The numerical time
step was chosen to be between $10^{-5}-5\cdot\nolinebreak10^{-4}$ (at a
typical shear rate of $\dot\gamma = 100$~Hz, the lower limit of the numerical
time step correspond to real time step of $10^{-7}$~s). The smaller range of
time steps is chosen when high ligand densities or stiff bonds (large $\kk$)
are considered (to avoid too large update steps at large direct forces in
\eq{langevin-euler-units}).  Keeping the other parameters fixed and varying
the rates $\pi,\epsilon_0$ on a grid in double logarithmic scale, we can
determine the different types of leukocyte motion in an \emph{on-off state
diagram}.
\begin{figure}[t!]
  \begin{center}
    \resizebox{.6\linewidth}{!}{\includegraphics{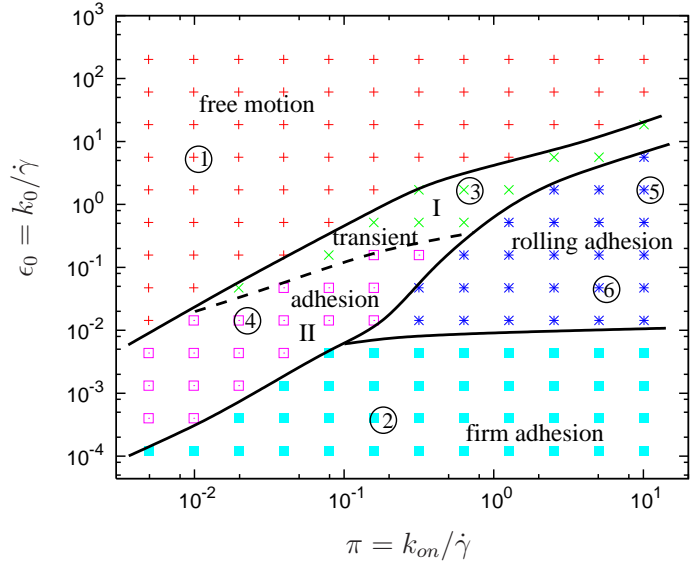}}
    \caption{(Color online) The on-off state diagram displays different
      stationary states of leukocyte motion obtained by simulations.  The
      ordinate shows the dimensionless unstressed off-rate $\epsilon_0$, the
      horizontal axis the dimensionless on-rate $\pi$.  Snapshots for the cell
      velocities at the $(\pi,\epsilon_0)$ parameter values marked by numbered
      circles are shown in \fig{fig:leuk:phaseexamples}.  Parameters used are
      the same as for \fig{fig:leuk:3dvonoff}.
      \label{fig:leuk:on-off-state}}
  \end{center}
\end{figure}

\fig{fig:leuk:on-off-state} shows a first example of such a state diagram.
The parameters used there are listed in the figure caption of
\fig{fig:leuk:3dvonoff} (as we keep the parameters $R,T_a, \Delta \rho, r_0$
fixed for all the diagrams shown in this section, they are not explicitly
listed for the following state diagrams).

All five states can be identified in \fig{fig:leuk:on-off-state} and in
\fig{fig:leuk:phaseexamples}(1-6) example trajectories for each of these
states are shown. In the limit of very large off-rates $\epsilon_0$ the cell
moves freely, i.\,e., no matter how frequently bonds are formed, force cannot
build up because dissociation occurs immediately. At very small off-rates the
cell is in the state of firm adhesion, i.\,e., once a bond is formed it lasts
long enough to stop the cell and to allow further bonds to be formed, thus
stabilizing the cell in its rest position.  In between these two limiting
cases for the off-rate we find the other three states.  From these the rolling
state appears only for on-rates $\pi$ above a certain threshold. This confirms
the conclusion drawn from early experiments with flow chambers that selectins
are especially suited for rolling adhesion due to their fast on- and off-rates
\cite{alon:97}. The state `transient I' appears when the off-rate is too large
to support rolling, but still too small to allow for free motion.  When both
on- and off-rates are small, the cell is in state `transient II'. In this
state the cell stops most likely whenever a bond is formed due to the small
off-rate.  This results in periods of firm adhesion. The small on-rate,
however, makes it rather unlikely that bonds are formed, which results in
periods of free motion. Thus, the cell moves in a stick-slip like fashion.

The dynamic state diagram \fig{fig:leuk:on-off-state} emphasizes that the
molecular rates (the on-rate $\pi$ and the off-rate $\epsilon_0$) are the main
determinants of rolling adhesion. In addition, the state diagram also depends
on the other parameters contained in the adhesive dynamics algorithm. In the
following we discuss the qualitative dependence on some of these parameters.

\begin{figure}[t!]
  \begin{center}
    \begin{tabular}{c@{\hspace{.04\linewidth}}c}
      \resizebox{.46\linewidth}{!}{\includegraphics{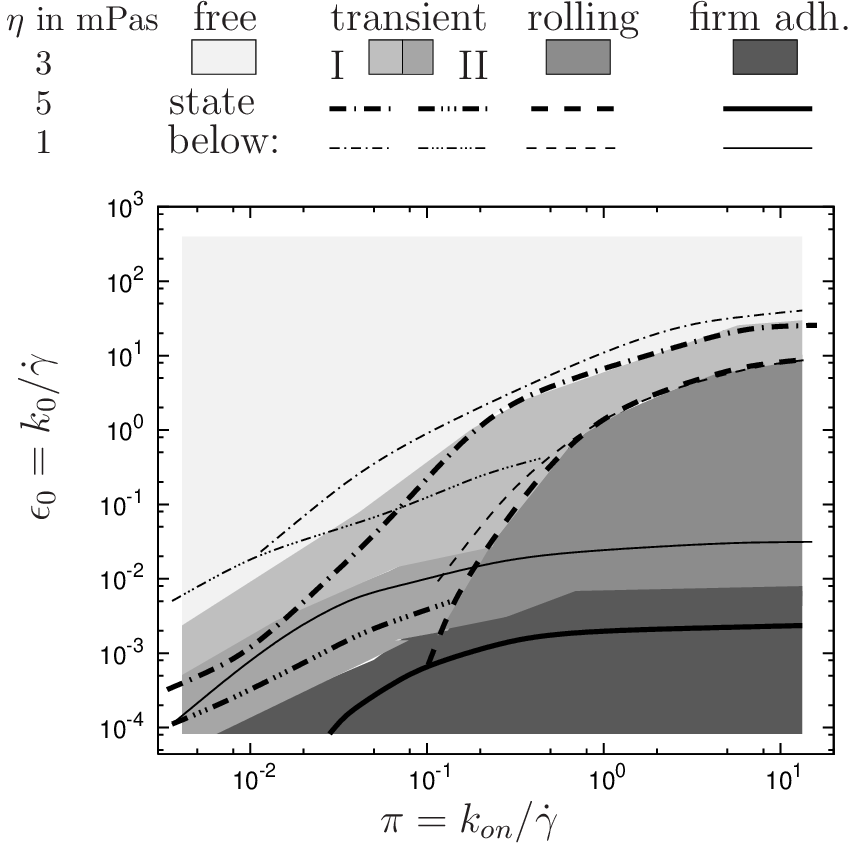}}&
      \resizebox{.46\linewidth}{!}{\includegraphics{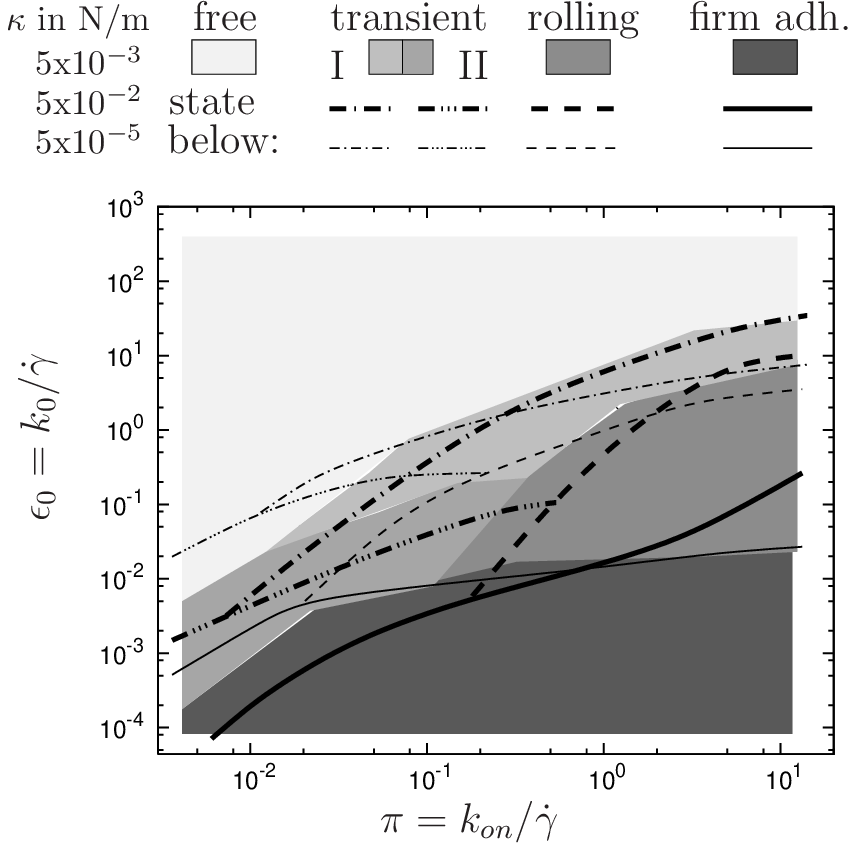}}\\
      (a) & (b) 
    \end{tabular}
    \caption{(a) On-off state diagrams for different viscosities $\eta$. The
      filled areas define the states of motion as determined from simulations
      at a reference viscosity $\eta = 3$~mPa\,s.  The lines denote the
      borders between the states of motion at a lower and at a higher viscosity
      than the reference viscosity.  The thin (thick) lines refer to a
      viscosity $\eta = 1$~mPa\,s (5~mPa\,s).  The figure legend tells which
      state is found below the corresponding line.  (b) States of motion at
      different spring constants $\kk$. The filled areas are the states of
      motion for the intermediate spring constant $\kk = 5\cdot
      10^{-3}$~N/m. The thin (thick) lines refer to $\kk = 5\cdot 10^{-5}$~N/m
      ($\kk = 5\cdot 10^{-2}$~N/m).  (Other parameters used for simulations in
      both (a) and (b): Number of receptors $N_r = 2500$, ligand-ligand
      distance $d = 2.5\cdot 10^{-2}~R$,
      in (a): $\kk = 5\cdot 10^{-3}$~N/m; in (b): $\eta =
      1$~mPa\,s.)
     \label{fig:leuk:stateshift01}}
\end{center}
\end{figure}

\subsection{Medium viscosity}

The impact of viscosity is illustrated in \fig{fig:leuk:stateshift01}a.
There, the on-off diagrams for the fluid viscosities $\eta = 1,3,5$~mPa\,s are
shown.  The viscosity of the standard medium in which leukocytes are usually
perfused through flow chambers is about 1~mPa\,s.  For $\eta = 3$~mPa\,s the
states of motion are distinguished by areas filled with different grey scales.
For $\eta = 1,5$~mPa\,s only the border lines between two states are
shown. The figure legend explains which state can be found below a given line.
For the sake of clarity, the original (12,13)-simulation grids for the three
diagrams are omitted. \fig{fig:leuk:stateshift01}a clearly shows that the main
effect of increasing the viscosity is the increase in the range where rolling
is possible. More precisely, the larger the viscosity, the lower the off-rate
$\epsilon_0$ at which firm adhesion sets in.  This effect results from the
Bell model for bond dissociation. The shear stress $\eta\dot\gamma$ and the
maximum force in a tether bond are proportional to the viscosity $\eta$ (the
maximum force is the force which holds the cell at rest, see
\fig{fig:leuk:stopp}).  Therefore, an increase in viscosity from $\eta^*$ to
$\eta$ increases the off-rate like
\begin{align*}
  \epsilon = \epsilon_0\left[\exp(F_{B}^*/F_d)\right]^{(\eta/\eta^*)},
\end{align*}
with $F_{B}^*$ the bond force at viscosity $\eta^*$. As a rough estimate for
the bond force we use $F_{B}^* \approx \|\boldv{F}_S\|/\cos\chi$ (see
\fig{fig:leuk:stopp} for the definition of $\boldv{F}_S, \chi$; for the angle
we estimate $\chi \approx 65^\circ$ \cite{alon:97}).  Then, for the parameter
values used here we have $F_{B}^*/F_d \approx 0.7$ at $\eta^* = 1$~mPa\,s.
Thus, if at some viscosity $\eta$ firm adhesion occurs for off-rates smaller
than a certain value $\epsilon_0^{firm}$, i.\,e., for $\epsilon_0 <
\epsilon_0^{firm}$, then, we expect firm adhesion for $\eta^*$ to exist at the
same rate $\epsilon$. For $\eta^*$ this rate is estimated to be
$\epsilon_0^{firm}\exp(F_{B}^*/F_d)^{\eta/\eta^*-1}$.  With $\eta/\eta^* = 5$,
we therefore expect rolling at $\eta^*$ for $\epsilon_0 >
\epsilon_0^{firm}(\eta^*) \approx 15\epsilon_0^{firm}(\eta = 5\eta^*)$. A
factor of roughly this order of magnitude between the off-rates at the border
between rolling and firm adhesion for $\eta = 5$~mPa\,s and $\eta^* =
1$~mPa\,s can also be read off from \fig{fig:leuk:stateshift01}a. On the other
hand no essential shift in the borderline between the states rolling and the
'transient I' can be spotted when the viscosity is changed. This is the case
as the referred line occurs at rather large $\epsilon_0$ values at which
$\epsilon_0$ appears to dominate over the force (viscosity) dependent part in
the Bell equation for the total off-rate $\epsilon$.  These estimates suggest
that the rolling state disappears at even smaller viscosities (one or two
magnitudes smaller than that of water) than used in
\fig{fig:leuk:stateshift01}a. This is indeed observed for simulations in this
viscosity range (data not shown).  Note however that in this rage, the
assumption of small Reynolds number might fail.

\subsection{Bond spring constant}

\fig{fig:leuk:stateshift01}b shows the state diagrams for three different
spring constants $\kk_{stiff}, \kk_{int},\kk_{soft} = 5\cdot
10^{-2},5\cdot10^{-3},5\cdot10^{-5}$~N/m, respectively.  The intermediate
spring constant $\kk_{int}$ is of the same order of magnitude as the spring
constant of the bond of P-selectin and its ligand \cite{fritz:98}. The softest spring
constant $\kk_{soft}$ mimics the effect of soft microvilli \cite{shao:98}.
From \fig{fig:leuk:stateshift01}b we first notice that the firm adhesion state
for $\kk_{stiff}$ occurs at smaller off-rates $\epsilon_0$ for small on-rates
$\pi$ and at larger off-rates for large on-rates compared to the case of the
intermediate spring constant $\kk_{int}$.  A closer view identifies two
effects that are responsible for this observation at small on-rates. First,
the stiffness of the bond results in a small elongation which then leads to an
obtuse bond-wall angle (the angle $\chi$ in \fig{fig:leuk:stopp}). The more
obtuse this angle is, the more the bond must be loaded to compensate the shear
force. In addition, the transport of the cell and thus also the bond extension
is governed by the shear rate $\dot\gamma$. A stiffer bond is therefore loaded
faster.  Both the faster loading \cite{erdmann:04a} and the larger bond force
result in an effective increase of the off-rate $\epsilon$, which at small
on-rates can be compensated by smaller off-rates $\epsilon_0$. In contrast, at
larger on-rates the fast dissociation is compensated by fast binding of new
bonds and rebinding of just dissociated bonds. Therefore, the increase in the
effective $\epsilon$ plays a minor role. On the other hand the faster loading
leads to a faster stop of the cell, so that to maintain rolling even larger
rates $\epsilon_0$ are necessary in this range of $\pi$.

When we compare the intermediate spring constant $\kk_{int}$ with the soft one
$\kk_{soft}$, we see that the region of rolling shrinks for the soft case. In
the case of the soft spring constant at larger $\pi$ rolling turns into
'transient I' already at smaller off-rates $\epsilon_0$. In addition, rolling
occurs still at much smaller on-rates than in the case of $\kk_{int}$ and
$\kk_{stiff}$, respectively. Both of these observations have their origin in
the larger tether elongation that is possible for soft bonds. This elongation
effectively increases the contact area (i.\,e., the area on the cell surface
that is less than the capture length $r_0$ away from the wall). More
precisely, bonds can still only be formed within the contact region but can
then exist also outside the contact region. If the off-rate $\epsilon_0$ is
not too large, this effect leads to an effective increase in the number of
available receptors, which explains the rolling in the region of smaller
$\pi$. For larger $\epsilon_0$ bonds rupture quickly also at small bond
forces. In the case of soft bonds the bond force is then not sufficiently high
as to reduce the translational velocity to induce rolling. The state of motion
then rather appears to be 'transient I' or free motion.

\begin{figure}[t!]
  \begin{center}
    \begin{tabular}{c@{\hspace{.04\linewidth}}c}
      \resizebox{.46\linewidth}{!}{\includegraphics{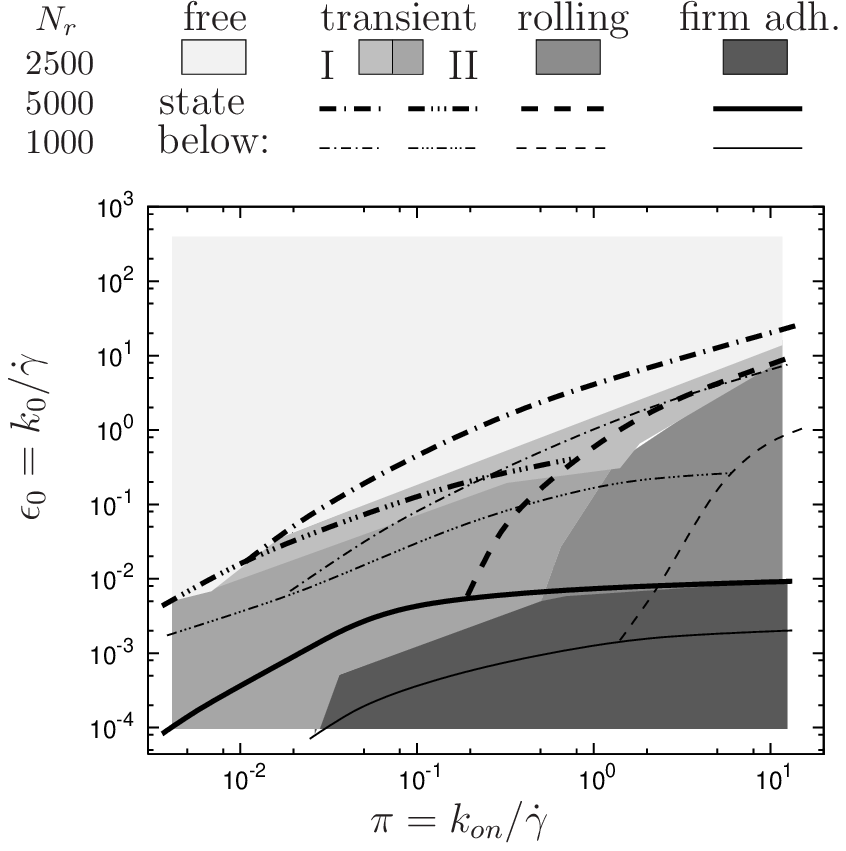}}&
      \resizebox{.46\linewidth}{!}{\includegraphics{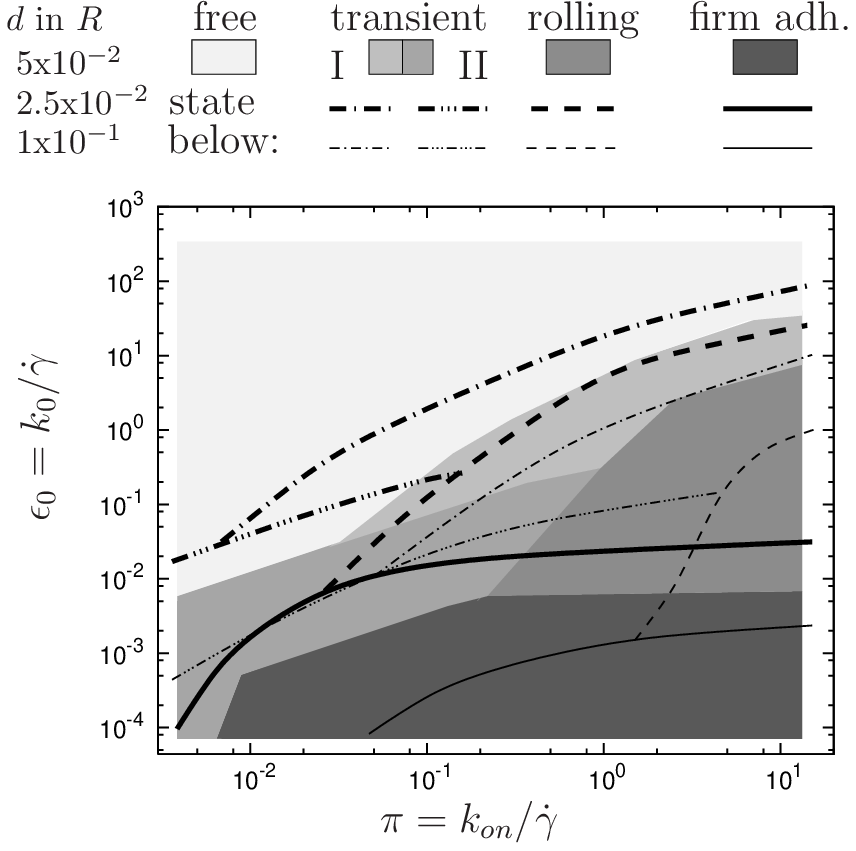}}\\
      (a) & (b) 
    \end{tabular}
    \caption{(a) On-off state diagrams for different numbers of
      receptors $N_r$. $N_r = 1000$ (thin lines) $N_r = 2500$ (filled
      areas) and $N_r = 5000$ (thick lines), ligand-ligand distance $d =
      5\cdot 10^{-2}$, $\kk =
      10^{-3}$~N/m, $\eta = 1$~mPa\,s.  (b) State diagrams for
      different ligand-ligand distances $d = 1\cdot 10^{-1}~R$ (thin lines) $d = 5\cdot 10^{-2}~R$
      (filled areas) and $d = 2.5\cdot 10^{-2}~R$ (thick lines), number of
      receptors $N_r = 5000$, $\kk = 5\cdot 10^{-3}$~N/m, $\eta = 1$~mPa\,s.  
     \label{fig:leuk:stateshift02}
    }
\end{center}
\end{figure}

\subsection{Receptor number}

Before discussing the impact of the number of receptor patches on the cell
surface, we first estimate the number of receptor patches within the contact
zone given $N_r$. Supposing the cell touches the wall it immediately follows
from basic geometrical considerations that on average $N_{r,contact} =
N_r~r_0/2$ receptor patches are in principle in capture range to wall ligands.
On the other hand, the average number of ligands in range to receptors is
calculated form the product of the projected area of the contact zone to the
wall and the ligand density. The radius of the projected contact area is
$\approx \sqrt{2Rr_0}$ as $r_0 \ll R$. Thus the number of ligands in the
contact zone is $N_{l, contact} = 2\pi \R r_0/d^2$.

In \fig{fig:leuk:stateshift02}a we have $N_r = 1000,2500,5000$ and $d =
5\cdot10^{-2}\R$, i.\,e., in the contact zone $N_{r,contact} \approx 5, 12.5,
25$ and $N_{l, contact} \approx 25$ (for $r_0 = 0.01R$). Therefore, the number
of receptors limits the maximum number of bonds in all three cases. In
addition, using as an estimate for the receptor patch density on the sphere
$\rho_r = \pi r_0^2 N_r/(4\pi R^2)$, we have $\rho_r \leq 0.125$, i.\,e., even
for $N_r = 5000$ receptor patches, they do not cover the contact zone
completely. So, not every receptor necessarily encounters a ligand, and
therefore the number of tether bonds is even less than $N_{r,contact}$.  In
fact, in the rolling state, we actually measure an average number of
$(1.3-2.6)$ for $N_r = 1000$, $(2-5)$ for $N_r = 2500$, and $(2-6)$ for $N_r =
5000$ existing bonds, respectively, depending on the actual on- and off-rate,
which is less than the respective $N_{r,contact}$.
\fig{fig:leuk:stateshift02}a shows that the larger $N_r$, the smaller on-rates
$\pi$ are sufficient to support rolling. Not every receptor-ligand encounter
(i.\,e., overlap of the receptor patch with a ligand) turns into a bond. This
happens only with a probability depending on $\pi$ and the dwelling time of
the encounter. Thus, the more encounter occur per time the more bonds will be
formed at a given rate $\pi$. As the encounter rate increases with increasing
number of receptor patches \cite{korn:06}, this also increases the average
number of bonds.

\fig{fig:leuk:stateshift02}a shows that the 'transient II' region expands with
decreasing $N_r$. The main effect here is that for smaller $N_r$ the rolling
turns into transient motion, while the border line between free and transient
motion is much less effected by the decrease in the number of receptor
patches.  The large $(\pi,\epsilon_0)$-range for the transient states is a
signature of few bonds being at work as we will see again when discussing the
influence of ligand-ligand distance. Then, single tethers slow the cell down
(depending on the off-rate they may either arrest the cell some while,
resulting in state `transient II', or just decelerate them resulting in state
`transient I'), but after dissociation it is unlikely that the current state
of motion is supported by further bonds.  However, as long as at least two
bonds are possible this effect is partly compensated at large on-rates when
the probability for receptor-ligand encounter to result in a bond is high.

\subsection{Ligand density}

To demonstrate the impact of ligand-ligand distance,
\fig{fig:leuk:stateshift02}b shows the state diagrams for $d = 10\cdot
10^{-2}~\R, 5\cdot 10^{-2}~\R, 2.5\cdot 10^{-2}~\R$ and $N_r = 5000$.  Using
the expressions derived in the previous subsection we now have $N_{l, contact}
\approx 6, 25, 100$, respectively, for the number of ligands in the contact
zone. The mean number of receptors that may form a bonds is $N_{r,contact} =
25$.  So, in principle $N_{r,contact}$ limits the number of bonds in the two
cases of higher ligand density and $N_{l, contact}$ is the limiting value only
for the very low ligand density. When we measure the average number of bonds
at high on-rates and relatively small off-rates (i.\,e. in the rolling and
firm adhesion region) we find for the intermediate ligand density with $d =
5\cdot 10^{-2}~\R$ about five bonds, and for the large ligand density with $d
= 2.5\cdot 10^{-2}~\R$ about fourteen bonds on average. The later value is
slightly larger than $N_{r,contact} = 12.5$. This can be explained with the
elasticity of the bonds that---once formed---allows them to also exist beyond
the contact zone. Thus only at very high ligand densities with $N_{l,
contact}\gg N_{r,contact}$, the number of receptors limits the maximum number
of bonds. For ligand densities with $N_{l, contact} \gtrsim N_{r,contact}$ a
decrease in $d$ still leads to an increase in the average number of bonds as
this increases the rate of receptor-ligand encounters.

The basic effect of decreasing the ligand density as illustrated by
\fig{fig:leuk:stateshift02}b is the shift of all state from the upper left
towards the lower right in the on-off state diagram. For example rolling is
only supported at larger on-rates and smaller off-rates when the ligand
density is decreased (i.\,e., $d$ is increased). Also the border line between
the state of free and transient motion is notedly shifted between the two
extreme cases of ligand density simulated for \fig{fig:leuk:stateshift02}b. In
fact, the shift of this border line is much more pronounced than in the
previous discussed case where the number of receptor patches was reduced. The
simple reason for that is that in \fig{fig:leuk:stateshift02}a $N_{r,contact}$
is changed by a factor of five whereas in \fig{fig:leuk:stateshift02}b the
$N_{l,contact}$ is changed by a factor of almost twenty.

Similarly to the case of reduced number of receptor patches, a
small ligand density results in an increased $(\pi,\epsilon_0)$ range
for transient adhesion. Thus, we note again that at a reduced rate of
receptor-ligand encounters, rolling tends to be converted to transient
motion.  We also note that if several parameters are changed at the
same time, the overall effect can be qualitatively understood by
superimposing the effects of the single parameter changes described
above (data not shown).

\subsection{Application: experimental determination of the on-rate}
\label{sec:application}

\begin{figure}[t!]
  \begin{center}
      \resizebox{.8\linewidth}{!}{\includegraphics{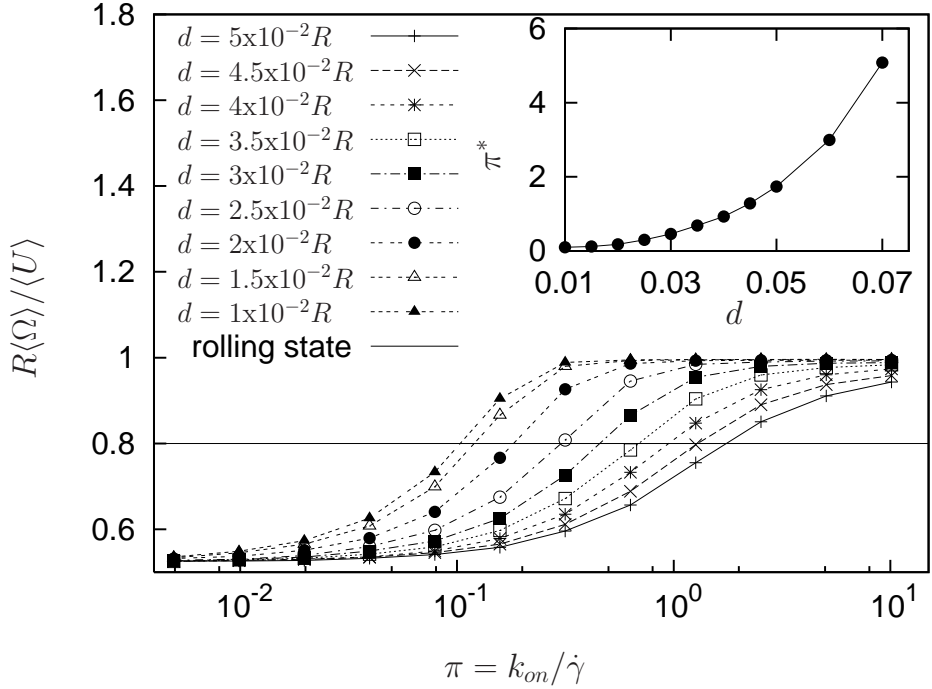}}
    \caption{Determination of the dynamic on-rate $\pi$ from the determination
    of the state of motion. At the fixed off-rate $\epsilon_0 = 0.16$, the
    ratio $R\mean{\Omega}/\mean{U}$ is plotted as a function of the on-rate
    $\pi$ for different ligand-to-ligand distances $d$ ranging from $d = 1\cdot
    10^{-2}$ to $d = 5\cdot 10^{2}$. In the $\pi$-range above the solid line
    at $R\mean{\Omega}/\mean{U} = 0.8$ the cell is rolling. In the inset the
    on-rate $\pi^*$ at which $R\mean{\Omega}/\mean{U} = 0.8$ is plotted as a
    function of $d$.  (Other parameters used: $N_r = 2500$, $d = 2.5\cdot
    10^{-2}~R$, $\dot\gamma = 100$~Hz, $\rc = 2\cdot10^{-11}$~m; $\kk = 5\cdot
    10^{-3}$~N/m; $\eta = 1$~mPa\,s.)
     \label{fig:pioffd}
    }
\end{center}
\end{figure}

The on-off state diagrams introduced here display the state of motion of a
cell or a receptor covered micro-bead for a given set of experimental
parameters in a wide range of possible on- and off-rates of the
receptor-ligand pair. Because the molecular properties of a given
receptor-ligand pair can be assumed to be given \emph{a priori}, each pair
corresponds to one point in the on-off state diagram. In the following we show
how the considerations regarding the rolling state presented here can be used
to determine the dynamic on-rate $\pi$ of a receptor-ligand pair by changing
the external parameter. For this line of reasoning the off-rate $\epsilon_0$
is assumed to be known, e.g.\ from dynamic force spectroscopy experiments.
\fig{fig:pioffd} shows the ratio of the mean velocities
$R\mean{\Omega}/\mean{U}$ of the cell as a function of $\pi$ for different
ligand-to-ligand distances $d$.  We see that for a given $d$ the cell is
slipping if the on-rate is small. Increasing the on-rate turns the cell's
motion into rolling, i.\,e., at some critical on-rate $\pi^*$ the ratio
$R\mean{\Omega}/\mean{U}$ reaches the value $0.8$ which we defined for
rolling. The smaller (larger) the ligand density (ligand-to-ligand distance
$d$) the larger is the critical on-rate $\pi^*$. More quantitatively this is
shown in the inset of \fig{fig:pioffd} where $\pi^*$ is plotted as a function
of $d$.  If one now changes the ligand density in a series of flow chamber
experiments from low to high values and by doing so determines the $d$-value
at which the motion of the cell turns from transient to rolling motion, then
one can read of from the inset of \fig{fig:pioffd} the on-rate $\pi$ of the
receptor-ligand bond.

The described procedure does not work for very small on-rates for which
rolling is not possible even for very large ligand densities (i.\,e., $d \ll
r_0$). On the other hand one can always find a ligand distance at which
rolling is not possible whatever value the on-rate takes. At very low ligand
densities $d\gg r_0$ the reaction is limited by the encounter rate and even
for very large rates $\pi$ rolling is not possible.

\section{Discussion}
\label{sec:discussion}

In this paper we introduced a new version of the adhesive dynamics
algorithm.  In contrast to earlier work, our approach also includes the
diffusive motion of cells resulting from thermal fluctuations.  This allowed
us to spatially resolve receptors and ligands. An immediate advantage of this
approach is that the single bond on-rate $k_{on}$ can be chosen to be
independent of the relative motion of cell and substrate. In this work, we
focused on the different dynamic states of motion which can be identified on
the cellular level. We first noticed that the action of a single bond not only
slows the cell down, but also changes the motion from slipping (which is the
case for cells moving free in hydrodynamic flow) to rolling in the sense
$\R\Omega/U \rightarrow 1$. In the case of multiple bonds, rolling can also be
observed regarding the mean values of the velocities, i.\,e.,
$\R\mean{\Omega}/\mean{U} \rightarrow 1$ at proper rates of bond formation and
rupturing. This motivated us to define the state of rolling adhesion as
$\R\mean{\Omega}/\mean{U} > 0.8$. By extending these observations we defined
five distinct states of stationary cell motion. These states were then
displayed in so-called on-off state diagrams which showed the impact of
different molecular rates. In addition, we investigated the effect of external
parameters.  For example, we showed that the cellular motion is changed
considerably when the viscosity of the medium is changed.

Our work shows that different dynamic states of cell motion can be defined in
a systematic and quantitative way. In particular, calculations of state
diagrams allow us to obtain a complete understanding of the way molecular and
other parameters determine motion on the scale of a cell. Similar approaches
have been taken before \cite{bruinsma:96,hammer:04a}, but without a proper
definition of rolling in the mechanical sense. Our computer work now shows
that including the rotational degrees of freedom allows us to investigate the
transition from hydrodynamic slipping to bond-mediated rolling in a more
detailed way.  Experimentally it is certainly a challenge to obtain similar
data in flow chamber experiments for micron-sized beads or cells. In
principle, making use of recent nanotechnological developments one could
attach receptors to micron-sized beads which are covered with anisotropic
surface layers \cite{burmeister:98,leiderer:05}. If these layers are
anisotropically reflective, rotational motion of the spheres can be recorded.
For cells, one would have to track surface or intracellular
markers (e.g.\ mitochondria or nuclei).

We also suggested a new procedure to experimentally determine values for
on-rates from monitoring cell motion (cf. \sec{sec:application}).  In contrast
to the off-rate, which can be measured e.g.\ from dynamic force
spectroscopy experiments, it is very difficult to infer values for the on-rate
$k_{on}$ in cell adhesion, where both partners have to be attached to
appropriate surfaces. Here, also one could make use of recent developments in
nanotechnology. As we have shown above, the ligand-to-ligand distance $d$ is a
crucial control parameter in our system.  Recently, is has become possible to
control this parameter using nanopatterend and biofunctionalized arrays of
gold dots \cite{spatz:04}. Therefore, in the future our predictions regarding
the effect of ligand positioning might be compared directly to experimental
data, especially if compared with a measurement of the rotational degrees of
freedom.

For conceptual and computational simplicity, here we have used the Bell model
for the force dependence of the off-rate $k_{off}$.  In principle, it would be
easy to include more complicated rupture scenarios, like the catch-slip
behavior recently reported for both P- and L-selectin
\cite{marshall:03,yago:04}. There is good reason to believe that this
molecular behavior is essential for the physiological function of these
molecules and different theoretical models have been suggested to explain this
behavior \cite{evans:04,thomas:06,pereverzev:06}. If combined with our
modeling framework for adhesive dynamics, these models might be tested against
experimental data from flow chamber experiments.  Such an approach would have
the big advantage that it avoids testing single molecules outside their
physiological environment, which is especially problematic for adhesion
receptors which usually are embedded in the plasma membrane and regulated by
the cytoskeleton.

Further possible extensions of our simulation framework include models
for cell deformability and hydrodynamic interactions between cells.
Cell deformations in free flow should become relevant only at shear
rates well above 100 Hz \cite{korn:07a}. In adhesion, cell deformation
depends also on the number and strength of adhesion bonds. In the case
of strong adhesion, it is well known that also viscoelastic changes
occur, including elongation of microvilli \cite{shao:98}. For
computational and conceptual simplicity, here we have focused on the
case of moderate shear flow and adhesion, when the effect of cell
deformability is small. Moreover deformations are irrelevant for rigid
microbeads, which have been shown to results in similar physical
effects as described for cellular systems
\cite{hammer:96a,hammer:00a,yago:07}. In order to combine elastic
effects and hydrodynamics, a very powerful framework is provided by
multi-particle collision dynamics \cite{gompper:05}, which in
principle also would allow to include non-laminar flow conditions and
hydrodynamic interactions between cells. However, these effects can be
safely avoided in flow chamber experiments by using sufficiently small
shear rates and cell numbers. As explained above, an elegant way to
test our theoretical predictions experimentally would be the
combination of appropriately coated microbeads with nano-structured
substrates.

Finally we comment on the applicability of our approach to other systems that
are based on the stochastic interplay between transport and adhesion. In the
work presented here, we have considered a situation where receptor and ligands
are both tethered to macroscopic surfaces. However, similar physical processes
are relevant if the two molecular binding partners are free in solution, for
example in affinity measurements \cite{wofsy:02} or shear-induced adhesion of
blood-clotting factors \cite{schneider:07}. Another biological system for
which the concepts for bond formation, bond rupture and transport discussed
here can be applied is the cargo transport by multiple molecular motors
\cite{klumpp:05,beeg:07}. Compared with rolling adhesion, the cell is replaced
by the cargo (e.\,g., a vesicle), the cell-anchored receptors by molecular
motors and the substrate-anchored ligands by binding sites on the filament. In
contrast to the case of rolling adhesion, now the system is not driven by some
external force, but the molecular motors actively step forward and pull their
cargo against an external viscous friction force. Our approach can also be
applied to non-biological systems. For example, it is well known that erratic
motion also occurs in the context of sliding friction. A flat slider that is
pulled above a plain wall exhibits stick-slip motion in some range of the
pulling velocity. It was shown recently that this special kind of motion can
be explained assuming the stochastic formation and rupture of molecular bonds
between the slider and the wall \cite{urbakh:04}. Thus the stick-slip motion
of sliding friction and the erratic movement of cells in rolling adhesions
seem to be based on the similar physical principles.

\begin{acknowledgments}
  This work was supported by the Center for Modeling and Simulation in
  the Biosciences (BIOMS) and the Cluster of Excellence Cell Networks
  at Heidelberg. Additional support came from the Minerva Foundation
  through a short term research grant for CBK.  We thank Ronen Alon
  and the members of his group for many interesting discussions on
  rolling adhesion.
\end{acknowledgments}

\appendix

\section{Bond dynamics algorithm}
\label{appendix:addyn}

The sphere's motion is described by \eq{langevin-euler-units}.  If no bond
between the sphere receptors and wall ligands exists, we take for the direct
force $\boldv{F}^D$ only gravity into account, i.\,e., the six-dimensional
force and torque vector is given by $\boldv{F}^D = (-\Delta m g \boldv{e}_z,
\boldv{0})$ with $g$ the earth acceleration constant. Receptor-ligand bonds
lead to additional contributions to both the force- and momentum-part of
$\boldv{F}^D$. More precisely, a bond between a ligand located at
$\boldv{r}_l$ and a receptor located at $\boldv{r}_r$ on the sphere's surface
(see \fig{fig:model:bdynamics}) pulls with a force
\begin{align}
  \label{eq:model:singlebondforce}
  \boldv{F}_B = \hat{\boldv{r}}_b F(r_b), \ha \hat{\boldv{r}}_b := \frac{\boldv{r}_{l} -
    \boldv{r}_{r}}{\|\boldv{r}_{l} - \boldv{r}_{r}\|}, \ha r_b :=  \|\boldv{r}_{l} - \boldv{r}_{r}\|.
\end{align}
$F(x)$ is the force extension curve that describes by what force the bond must 
be pulled to stretch it up to a total length $x$. 
Here, we consider the bonds to be semi-harmonic springs (cable model)
\begin{align}
  \label{eq:model:cable}
  F(x) = \kk (x - l_0) \Theta(x-l_0),\ha  \Theta(x) := \left \{ 
  \begin{array}{l}1,\ha x > 0\\0,\ha \mbox{else} \end{array}
  \right., 
\end{align}
with $l_0$ the resting length and $\kk$ the spring constant.  The cable model
is the simplest model for polymeric tethers.  In the cable model a bond
behaves as a spring only if it is stretched (extension larger than the resting
length), otherwise the bond exerts no force on the sphere.  Treating the
receptor-ligand complex as a harmonic spring works fine in the small extension
regime \cite{fritz:98}. For large extensions the force extension curve for
polymers is supposed to grow much faster than linear, and when the bond
extension approaches the total contour length of the receptor-ligand complex
it even diverges (strain stiffening). However, typical bonds are weak and
their rupture probability increases exponentially with force. Therefore, we
expect bond extensions to be restricted to the linear regime.  As the bond
force pulls on the sphere's surface also a torque results
\begin{align*}
     \boldv{T}_B = \hat{\boldv{r}}
     \times \boldv{F}_B(\boldv{r}_b), 
\end{align*}
where $\hat {\boldv{r}}$ is the connection vector from the center of
the sphere to the point on its surface where the receptor is attached
(see \fig{fig:leuk:stopp}).  Thus, the total force and torque
contribution to $\boldv{F}^D$ by the bonds is
\begin{align} 
  \label{totforce}
  \kk\sum_{i = 1}^{N_{r}}q_i F(r^i_b)\left(\hat{\boldv{r}}^i_b,
    \hat{\boldv{r}}^i \times \hat{\boldv{r}}^i_b
  \right),
\end{align}
with $N_r$ the total number of receptors and $q_i = 1$ if the $i$th receptor
forms a bond and zero otherwise. The $q_i, i = 1,\ldots,N_r$ are stochastic
variables. Thus, the contribution \eq{totforce} lets the direct force
$\boldv{F}^D$ also become a stochastic variable.

With this at hand we can now define the \emph{adhesive dynamics} rules,
applied in each update step $\Delta t$ (cf. \fig{fig:model:bdynamics}):
\begin{itemize}
  \item[(i)] The sphere's position and orientation is updated according to
    \eq{langevin-euler-units} (for an explicit description see
    Ref.~\cite{korn:07a}).
  \item[(ii)] The receptor positions in the flow chamber coordinate system
  are calculated.
  \item[(iii)] Each inactive receptor is represented by a capture ball with
  radius $r_0 \ll 1$.
  \item[(iv)] If the distance between a receptor and any ligand is $\leq
    r_0$ a bond is established with probability $p_{on} = 1 -
    \exp(-\Delta t\cdot k_{on})$, then the resting length of the bond
    is set to the receptor-ligand distance at the instance of
    bond-formation (i.\,e., the bond force at the moment of bond
    formation is zero)  and is stored together with the ligand
    position.  A bond can only be formed if the corresponding receptor
    and ligand are not already part of another bond.
  \item[(v)] For each active bond, the contribution to $\boldv{F}^D$ is
  calculated.
  \item[(vi)] Each existing bond dissociates with a rate given by the
    Bell equation, \eq{eq:model:bell}.  Thus, each bond ruptures with
    probability $p_{off} = 1-\exp(-\Delta t\cdot k_{off}(F))$, where
    $F$ is the instantaneous force acting along this bond.
\end{itemize}
When a bond has ruptured, both the receptor and the ligand can form a new bond
in the next time step according to the rule (iv). As for the resting length
$l_0$ of a bond always $l_0 < r_0 \ll 1$ is true, modeling bonds as harmonic
springs in both the extension and compression regime would not make much
difference to the results that are obtained by the cable model. Given the
probability for bond formation or rupture $p_{on}$ or $p_{off}$, respectively,
a standard Monte-Carlo technique is used to decide whether the action
happens or not: Using a pseudo-random number generator a random number $rand$
from the uniform distribution in the interval $[0,1]$ is drawn. If then
$p_{on/off} > rand$ the respective action takes place, otherwise not.


\end{document}